\documentclass[aps,prl,twocolumn,superscriptaddress,showpacs,floatfix]{revtex4-2}

\usepackage{float}
\usepackage{graphicx}
\usepackage{amsmath}
\usepackage{amssymb}
\usepackage{hyperref}
\usepackage{xcolor}
\usepackage{bm}
\usepackage{cleveref}
\crefname{figure}{Fig.}{Figs.}
\Crefname{figure}{Figure}{Figures}
\crefname{table}{Table}{Tables}
\Crefname{table}{Table}{Tables}
\crefformat{equation}{Eq.~(#2#1#3)}
\Crefformat{equation}{Equation~(#2#1#3)}

\begin{document}
\rightline{FERMILAB-PUB-26-0558-T}
\title{Measurement of the Hubble constant with high-energy neutrinos}

\author{Gonzalo~Herrera}
\email{gonzaloh@mit.edu}
\affiliation{Department of Physics and Kavli Institute for Astrophysics and Space Research, Massachusetts Institute of Technology, Cambridge, MA 02139, USA}
\affiliation{Harvard University, Department of Physics and Laboratory for Particle Physics and Cosmology, Cambridge, MA 02138, USA}

\author{Nicholas~Kamp}
\email{nkamp@fas.harvard.edu}
\affiliation{Harvard University, Department of Physics and Laboratory for Particle Physics and Cosmology, Cambridge, MA 02138, USA}

\author{Carlos~A.~Arg\"{u}elles}
\email{carguelles@fas.harvard.edu}
\affiliation{Harvard University, Department of Physics and Laboratory for Particle Physics and Cosmology, Cambridge, MA 02138, USA}

\begin{abstract}
Measuring distances in the Universe is one of the hardest problems in physics and astronomy. Almost every distance probe relies on photons, whose propagation across cosmic distances introduces extinction, absorption, scattering, and radiative-transfer effects. Neutrinos suffer none of these and propagate unattenuated through dust, intergalactic medium, and dense source environments alike. We introduce a new distance-ladder method for measuring the Hubble constant $H_0$ using high-energy astrophysical neutrinos from point sources as standardizable candles, and report its first observational realization. Using 12 X-ray--selected Seyfert galaxies for which IceCube reports significant per-source neutrino excesses in its 14-year public point-source release, we exploit the disk-corona correlation $L_\nu = \kappa\, L_X^\beta$ between neutrino and X-ray luminosities to construct a neutrino distance ladder anchored by non-redshift distances to NGC~1068 (Cepheid + TRGB). We find $H_0 = 49^{+40}_{-30}\,\mathrm{km\,s^{-1}\,Mpc^{-1}}$ and $\beta = 0.67^{+0.16}_{-0.25}$ (68\% credible intervals), with the corona slope disfavoring the calorimetric limit $\beta = 1$ at ${\sim}2\sigma$. The result is consistent with existing $H_0$ determinations from Planck and SH0ES within 1$\sigma$. While the uncertainty on $H_0$ is large, the measurement is free of electromagnetic propagation systematics and demonstrates the viability of neutrinos as a novel cosmographical probe.
\end{abstract}

\maketitle


\textit{Introduction.}---Nearly everything we know about the scale of the
Universe arrives encoded in light. With the sole exception of
gravitational-wave sirens, every rung of the cosmic distance ladder
above parallax; variable stars, Type~Ia supernovae, baryon acoustic
oscillations, is read out in photons. Each photon-based rung therefore
inherits the same propagation effects: dust extinction, intergalactic
absorption, scattering, and the radiative-transfer modeling needed to convert
observed fluxes into intrinsic luminosities. The stakes of these shared
systematics are now plain. The two flagship determinations of the Hubble
constant disagree at the $\sim\!5\sigma$ level:
$H_0 = 67.4 \pm 0.5\,\mathrm{km\,s^{-1}\,Mpc^{-1}}$ from the cosmic microwave
background~\cite{Planck2018} versus
$73.0 \pm 1.0\,\mathrm{km\,s^{-1}\,Mpc^{-1}}$ from the Cepheid--supernova
ladder~\cite{SH0ES2022}; the Hubble tension. A distance measurement carried by a different
messenger would bypass the electromagnetic channel and its propagation systematics entirely.

High-energy astrophysical neutrinos are such a messenger.  They
traverse the Universe untouched by the dust, gas, and radiation
fields that degrade light: no extinction, no absorption, no
scattering.  The only propagation effect they undergo is flavor
oscillation, which redistributes neutrino types but leaves their
total number intact, and is well understood.  If the intrinsic
neutrino luminosity ($L_\nu$) of a source can be predicted from an
independent observable, neutrinos become standardizable candles, and
the distances inferred from their fluxes are free of every
electromagnetic propagation systematic that limits the existing rungs of the ladder. Here we realize this idea and report the first measurement of the
Hubble constant with neutrinos, finding
$H_0 = 49^{+40}_{-30}\,\mathrm{km\,s^{-1}\,Mpc^{-1}}$. Two observations make it possible.
\begin{figure}[H]
  \centering
  \includegraphics[width=\columnwidth]{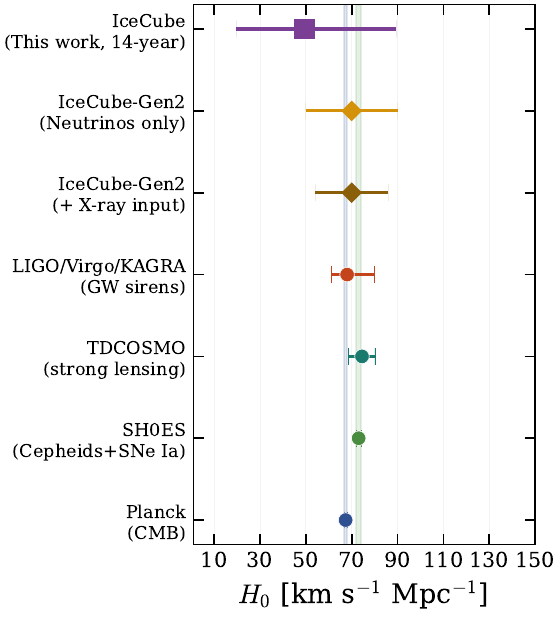}
  \caption{Comparison of our $H_0$ measurement (purple square) with
    the Planck CMB~\cite{Planck2018}, SH0ES
    Cepheid--supernova~\cite{SH0ES2022}, TDCOSMO strong
    lensing~\cite{TDCOSMO2020}, and LIGO/Virgo/KAGRA
    standard-siren~\cite{LIGO_GW_H0_2023} determinations, together
    with two projected IceCube-Gen2~\cite{IceCubeGen2}
    sensitivities (gold diamonds, drawn at $H_0 = 70$ for
    illustration).  \emph{Light gold}: a Seyfert-only Gen2 sample
    ($50$ sources, $8\times$ effective area, $25\%$ X-ray flux
    uncertainty, $10$ geometric anchors, $\kappa$ profiled).
    \emph{Gold}: adding input from X-ray modeling, i.e. fixing $\kappa$ and imposing an external Gaussian prior on the slope, $\sigma_\beta =0.02$).  Both forecasts use the Fisher formalism of the
    Supplemental Material~\cite{supplement}.
    Shaded bands indicate the Planck and SH0ES $1\sigma$ regions.}
  \label{fig:H0_whisker}
\end{figure}
First, the IceCube Neutrino Observatory has detected high-energy
neutrinos from individual active galactic nuclei: the nearby Seyfert
galaxy NGC~1068 at $4.2\sigma$~\cite{IceCube_NGC1068_2022}, and a
$3.3\sigma$ collective excess from X-ray--selected Seyferts in a
13-year population analysis~\cite{IceCube_Xray_2025}. Second, these
detections match disk-corona models in which the neutrino and X-ray
luminosities track each other through a power-law relation,
$L_\nu \propto L_X^{\beta}$~\cite{Stecker1991,Stecker2005,MKM2020,Kheirandish2021}. This
property makes the sources standardizable.  Anchoring this
correlation with a non-redshift distance to NGC~1068~\cite{NGC1068_Cepheid,NGC1068_TRGB}, we measure
$\beta$ directly from IceCube data and obtain $H_0$ from twelve
Seyfert galaxies.  The
result is a new rung of the cosmic distance ladder in which photons
enter only at the local anchor, never along the path.  Full
methodological details are provided in the Supplemental
Material~\cite{supplement}.

\textit{Model.}---Seyfert galaxies are expected to emit X-rays and
high-energy neutrinos from the same place: the corona, a compact
region of hot magnetised plasma surrounding the supermassive black
hole.  The corona's hot electrons upscatter accretion-disk photons
into X-rays, while protons accelerated in the same plasma collide
with the coronal gas ($pp$) and radiation ($p\gamma$) to produce
neutrinos~\cite{Stecker1991,Stecker2005,MKM2020,Kheirandish2021}.  Because both signals draw
on the same reservoir, their luminosities are expected to track one
another through a power-law correlation,
\begin{equation}
  L_\nu = \kappa\, L_X^{\,\beta},
  \label{eq:LnuLX}
\end{equation}
where $L_X$ is the X-ray luminosity in the Swift-BAT band
(20--50\,keV), $L_\nu$ is the neutrino luminosity in the
IceCube-sensitive TeV--PeV band; observational definitions set by
the surveys and detectors used~\cite{supplement}, and the normalisation
$\kappa$ absorbs the bolometric-to-band conversions on each side.

The slope $\beta$ encodes the corona physics.  A value of
$\beta = 1$ corresponds to the calorimetric limit, in which pion
production is saturated and the proton-injection power scales
linearly with the X-ray luminosity~\cite{MKM2020}.  A slope $\beta > 1$ is the signature of thin-target $p\gamma$
production, where a brighter corona supplies more target photons and
hence more pion-production
opportunities~\cite{Stecker1991,Kheirandish2021}.  A slope $\beta < 1$
instead requires the proton-acceleration power to grow more slowly than
the X-ray output, or the neutrino yield per pion to be actively
suppressed at higher $L_X$.  Two mechanisms can produce such suppression
in brighter, denser coronae:
(i) synchrotron cooling of secondary pions and muons in the strong
coronal magnetic fields expected at high accretion rate
~\cite{WaxmanBahcall1997,MKM2020}, which truncates the meson decay chain
before it produces high-energy neutrinos;
and (ii) Bethe--Heitler pair production ($p\gamma \to p e^+ e^-$),
which diverts a fraction of the proton energy into $e^{\pm}$ pairs
rather than pions when the target photon field is sufficiently
dense~\cite{Inoue2019}.  Both scale with quantities that grow with
accretion power---magnetic-field strength $B$ and target photon density,
respectively---so both predict a sub-linear $L_\nu$--$L_X$ scaling in
the direction preferred by the data.  Magnetically powered corona models
provide a natural framework in which both effects operate
simultaneously~\cite{Kheirandish2021,Inoue2019}.  Regardless
of the regime, \cref{eq:LnuLX} is universal across the population to
the extent that the sources share comparable coronal spectral shapes
within these bands; an assumption supported by the observed
spectral homogeneity of BAT-selected Seyferts; residual
source-to-source variation is absorbed into the per-source
measurement uncertainty~\cite{supplement}.

A non-trivial slope $\beta \neq 1$ turns the $L_\nu$--$L_X$
correlation into a distance probe.  Combining the inverse-square law
$L = 4\pi d_L^2 F$ with \cref{eq:LnuLX} gives
$F_\nu \propto F_X^{\,\beta}\, d_L^{\,2(\beta-1)}$: two sources with
the same X-ray flux but different distances predict different
neutrino fluxes, and the mismatch measures the distance.  When
$\beta = 1$ exactly, the distance dependence cancels and the method
carries no information about $H_0$.

A non-redshift distance to even one anchor galaxy breaks the circularity inherent in this
construction; without it, the same flux ratios would have to
determine both the normalisation $\kappa$ and $H_0$.  The primary analysis anchors the ladder to the single Cepheid+TRGB distance to NGC~1068~\cite{NGC1068_Cepheid,NGC1068_TRGB} ($10.93 \pm 0.74\,\mathrm{Mpc}$), which fixes the absolute scale of the correlation.  The remaining eleven sources lie in the
Hubble flow (out to $z \approx 0.056$), where their luminosity
distances depend on $H_0$; it is this dependence that the data
exploit to constrain $H_0$.

Energy redshifting and spectral index corrections are negligible across this sample.  A dedicated
sensitivity run bounds the correction below $\sim\!5\%$ for $z < 0.06$ and shifts
$H_0$ by less than $2\,\mathrm{km\,s^{-1}\,Mpc^{-1}}$~\cite{supplement}.

\textit{Likelihood.}---We fit the correlation in luminosity space.
The corona-model relation $L_\nu = \kappa\, L_X^{\,\beta}$ holds
between rest-frame, band-integrated luminosities, so both
sides of the data must be converted into them.  The neutrino side
starts from the IceCube signal counts $\hat{n}_{s,i}$, extracted
from a profile-likelihood scan of the 14-year public point-source
data with the SkyLLH unbinned likelihood~\cite{IceCube_14yr,SkyLLH};
the X-ray side starts from the observed fluxes $F_{X,i}$.  These are
converted to rest-frame luminosities as follows.  The neutrino signal
counts are converted to a band-integrated luminosity
$\hat{L}_{\nu,i} = 4\pi d_{L,i}^{\,2}\,\hat{n}_{s,i}/C_i$ using a
per-source IceCube detector acceptance~\cite{supplement}, which
encodes the effective area at the source declination and the assumed
neutrino spectral index; the X-ray fluxes are converted to X-ray
luminosities via $L_{X,i} = 4\pi d_{L,i}^{\,2}\,F_{X,i}$.
The luminosity distance is $d_L(z, H_0)$ in flat $\Lambda$CDM with
$\Omega_m = 0.315$ for the Hubble-flow sources, or the corresponding
anchor distance for the calibrator.  Model and data are
then compared through a Gaussian likelihood
\begin{equation}
  \ln\mathcal{L} = \sum_i \left[
    -\,\frac{1}{2}\,\frac{\big(\hat{L}_{\nu,i} - L_{\nu,i}^{\,\rm pred}\big)^2}
         {\sigma_{L,i}^2}
    \;-\; \ln\sigma_{n,i} \right]\,,
  \label{eq:lnL}
\end{equation}
with $L_{\nu,i}^{\rm pred} = \kappa\, L_{X,i}^{\,\beta}$ and
$\sigma_{L,i}^2$ combining the propagated SkyLLH measurement
uncertainty with the X-ray flux error; the normalization carries
the corresponding count-space width $\sigma_{n,i}$, the Jacobian
of the counts-to-luminosity conversion having cancelled its
distance dependence~\cite{supplement}.

The Hubble constant enters \cref{eq:lnL} only through the
$d_L^{\,2(\beta-1)}$ lever arm identified above.  The distance
appears in both terms of each residual
$\hat{L}_{\nu,i}-L_{\nu,i}^{\rm pred}$; as $d_L^{\,2}$ in the
measured luminosity and as $d_L^{\,2\beta}$ in the prediction, so
its leading effect cancels and the sensitivity to $H_0$ is finite
whenever $\beta \neq 1$.

The overall normalisation $\kappa$ is profiled out by iteratively reweighted least squares~\cite{supplement}; maximised over, rather than sampled, at each likelihood evaluation, leaving a three-dimensional parameter space
$\{H_0, \beta, d_{1068}\}$, where the non-redshift distance prior on $d_{1068}$ covers
NGC~1068 (Cepheid + tip-of-the-red-giant-branch envelope)~\cite{NGC1068_Cepheid,NGC1068_TRGB}; sensitivity to adding NGC~4151 and NGC~7469 as secondary anchors is presented in the Supplemental Material~\cite{supplement}.  The $25\%$ X-ray flux uncertainty (multiplicative on the
prediction, and conservatively taken) already absorbs any intrinsic scatter in the
$L_\nu$--$L_X$ relation; the two enter the likelihood identically
and are degenerate, so we do not fit a separate scatter parameter,
which with only 12 sources would be unconstrained.

\textit{Distance ladder.}---The absolute calibration of the ladder
rests on the single non-redshift distance to NGC~1068,
$10.93 \pm 0.74\,\mathrm{Mpc}$ (Cepheid + tip-of-the-red-giant-branch
envelope~\cite{NGC1068_Cepheid,NGC1068_TRGB}).  We choose NGC~1068 as the sole primary
anchor because it plays, for this ladder, the role the LMC plays for
SH0ES~\cite{SH0ES2022}: it is the closest galaxy in the sample, and its
proximity is what makes both individual Cepheid variables and
tip-of-the-red-giant-branch stars simultaneously resolvable, yielding the
Cepheid+TRGB envelope with a $6.8\%$ fractional uncertainty; smaller than
any other geometric distance in the sample.  It is also the only
individually significant ($4.2\sigma$) neutrino source in our sample
\cite{IceCube_NGC1068_2022}, and the archetype obscured Seyfert-2 in which
the disk-corona neutrino production model was originally developed and
tested~\cite{MKM2020,Kheirandish2021,Inoue2019}.  The anchor enters as a
Gaussian prior on the sampled parameter $d_{1068}$, so its uncertainty
propagates fully into the $H_0$ posterior.  One-, two-, and three-anchor
variants are examined in the Supplemental Material~\cite{supplement}.

\begin{figure*}[t!]
  \centering
  \includegraphics[width=2.0\columnwidth]{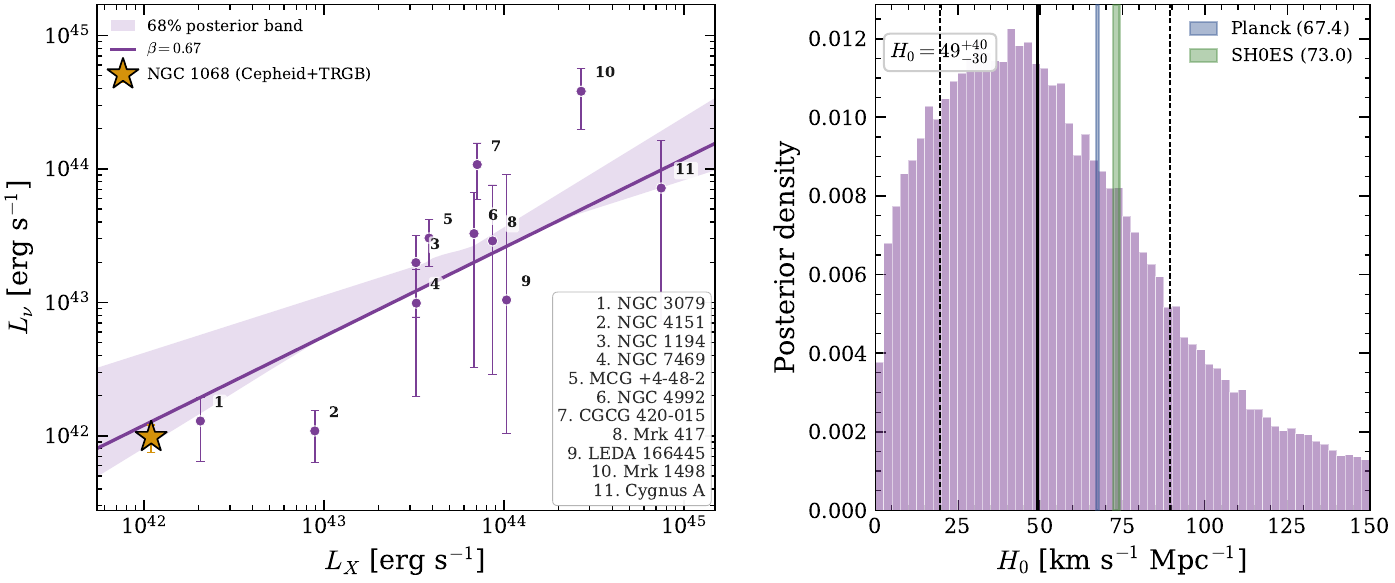}
  \caption{\textbf{Left:} $L_\nu$--$L_X$ relation for the eleven
    Hubble-flow Seyferts (blue circles) and the single non-redshift-distance
    calibrator NGC~1068 (gold star, Cepheid+TRGB envelope), evaluated at
    the posterior median $H_0 = 49\,\mathrm{km\,s^{-1}\,Mpc^{-1}}$.
    The solid line and shaded band show the best-fit power-law
    $L_\nu \propto L_X^{\,\beta}$ with $\beta = 0.67$ and the
    68\% credible band on the slope from the joint posterior.
    \textbf{Right:} marginal posterior distribution of the Hubble
    constant $H_0$.  The solid black line marks the median and
    dashed lines bracket the 68\% credible interval; dotted lines
    indicate the Planck ($67.4$) and SH0ES ($73.0$) reference
    values in $\mathrm{km\,s^{-1}\,Mpc^{-1}}$.}
  \label{fig:posteriors}
\end{figure*}

\textit{Data.}---Our sample comprises the 12 Seyfert galaxies that
IceCube's 13-year population analysis~\cite{IceCube_Xray_2025}
identified as the X-ray--selected sources with the strongest
individual neutrino excesses (collective significance $3.3\sigma$;
see Supplemental Material Table~V~\cite{supplement} for the full
source list).  Their intrinsic 20--50\,keV X-ray fluxes
$F_X$ are taken verbatim from that reference, which
corrects for absorption using column-density fits from the BAT AGN
Spectroscopic Survey (BASS)~\cite{BASS,SwiftBAT}; the
20--50\,keV band is chosen precisely because it minimises this
correction. As a cross-check, repeating the entire analysis with intrinsic 2--10\,keV fluxes compiled from per-source torus-model X-ray analyses reproduces the baseline result when a low NGC~1068 intrinsic luminosity ($\sim 10^{42}$\,erg\,s$^{-1}$) is adopted \cite{Rodri}, while higher values favored in some neutrino-emission models ($4\times10^{43}$\,erg\,s$^{-1}$) \cite{Bauer}, moves the posteriors to $\beta = 0.40^{+0.22}_{-0.19}$ and $H_0 = 109^{+29}_{-37}$\,km\,s$^{-1}$\,Mpc$^{-1}$ (see Supplemental Material~\cite{supplement}).  We extract the neutrino measurements ourselves,
deriving signal counts and uncertainties from the publicly
available 14-year point-source data~\cite{IceCube_14yr} with the
SkyLLH framework~\cite{SkyLLH}.
Because the sample is selected on
neutrino significance, the selection itself can imprint an apparent
$L_\nu$--$L_X$ correlation~\cite{LnuLX_spurious2026}; the
Supplemental Material~\cite{supplement} discusses this bias and the
features of our likelihood that mitigate, but do not eliminate, it.
For each source we convert the reported heliocentric redshift to the
CMB rest frame following the Planck 2018 dipole~\cite{Planck2018,Carr2022},
subtract the measured peculiar velocity from the Cosmicflows-4
catalogue~\cite{Carrick2015} where a direct-distance measurement is
available (NGC~1068, NGC~4151, NGC~3079, NGC~7469), and include a
residual $\sigma_v = 250\,\mathrm{km\,s^{-1}}$ uncertainty per source
in the likelihood variance for the remaining Hubble-flow sources.
Both corrections are standard in the $H_0$ literature~\cite{SH0ES2022}
and are essential at our lowest redshifts ($z \approx 0.003$--$0.004$),
where the CMB-frame shift alone is $\sim 20$--$30\%$; the
Supplemental Material details their construction and impact.

\textit{Results.}---From the joint three-dimensional posterior over $(H_0,\beta,d_{\rm NGC\,1068})$ we
measure
\begin{align}
  H_0 &= 49^{+40}_{-30}\,\mathrm{km\,s^{-1}\,Mpc^{-1}}
  \quad\text{(68\% CI)}\,, \label{eq:H0}\\
  \beta &= 0.67^{+0.16}_{-0.25}
  \quad\text{(68\% CI)}\,, \label{eq:beta}
\end{align}
where CI denotes the credible interval.  \Cref{fig:posteriors}
shows the fitted $L_\nu$--$L_X$ relation for the full sample and
the marginal $H_0$ posterior.  The fractional uncertainty on $H_0$
is dominated by the per-source counting statistics of the eleven
Hubble-flow sources, not by the non-redshift
anchors~\cite{supplement}.  \Cref{fig:H0_whisker} places it
alongside the Planck CMB, SH0ES Cepheid--supernova, TDCOSMO
strong-lensing, and LIGO/Virgo/KAGRA standard-siren
determinations.

The data prefer a sub-linear slope: a hundredfold spread in $L_X$
translates into only a factor-$\sim\!22$ spread in $L_\nu$, below
the calorimetric prediction of $\beta = 1$.  The marginal
posterior excludes $\beta = 1$ at ${\sim}2.0\sigma$, and a Gaussian
prior centred on the calorimetric value is pulled down by the data
to $\beta = 0.86 \pm 0.08$~\cite{supplement}.  This is the first
joint measurement of $\beta$ across a population of individually
resolved Seyferts, and it indicates that their coronae are
subcalorimetric: protons may escape before losing their full energy to
collisions.

The present 12-source measurement does not yet weigh in on the
Hubble tension: the posterior is consistent with both the Planck
and SH0ES values~\cite{Planck2018,SH0ES2022} at the $1\sigma$
level.

IceCube-Gen2~\cite{IceCubeGen2} can sharpen the measurement
substantially, as illustrated in \cref{fig:H0_whisker}.  The forecast uses a Fisher information matrix~\cite{supplement},
with per-source signal-to-noise growing as
$\sqrt{A_{\rm eff,Gen2}/A_{\rm eff,IC}} \approx \sqrt{8}$ and the
cosmological leverage entering through the slope factor
$|2(1-\beta)|$.  We marginalize over the $L_\nu$--$L_X$ normalization
$\kappa$;
the $H_0$ information then comes entirely from the geometrically
anchored sources, which break the $\kappa$--$H_0$ degeneracy.
Evaluated at the measured slope $\beta = 0.67$, a Gen2 sample of
$\sim\!50$ Seyferts with $\sim\!10$ anchors (e.g., from TRGB or Cepheid distance measurements) gives
$\sigma(H_0) \approx 20\,\mathrm{km\,s^{-1}\,Mpc^{-1}}$.
This sensitivity can be improved to $\sigma(H_0) \approx 16\,\mathrm{km\,s^{-1}\,Mpc^{-1}}$ with additional input from X-ray modeling, i.e., if the normalization $\kappa$ in \cref{eq:LnuLX} can be fixed and $\beta$ can be constrained with an external prior uncertainty of $\sigma_\beta =0.02$~\cite{supplement}.

These projections hold the
X-ray flux uncertainty at a conservatively present $\sim 25\%$; intrinsic scatter in
the $L_\nu$--$L_X$ relation, if present, is degenerate with it and
absorbed into the same term.

The data themselves validate the assumed X-ray spectral shape:
rerunning the analysis with the photon index free returns
$\gamma_X = 2.0 \pm 0.7$, consistent with typical BAT-selected
Seyfert spectra, and confirms the
$<\!2\,\mathrm{km\,s^{-1}\,Mpc^{-1}}$ bound on the induced $H_0$
shift quoted above~\cite{supplement}.

\textit{Discussion.}---This is the first use of high-energy
neutrinos to measure cosmic distances.  The fractional
uncertainty on $H_0$ is larger than that of established probes,
but three features distinguish the approach.

First, neutrinos accumulate no systematics in transit.  Each
established probe carries its own residual propagation
uncertainty; dust extinction for Type~Ia supernovae, the
mass-sheet degeneracy for time-delay lensing, inclination for
gravitational-wave sirens, whereas for neutrinos the entire error
budget lives in the source physics: the $L_\nu$--$L_X$
correlation.

Second, our ladder mirrors the SH0ES Cepheid--supernova
ladder~\cite{SH0ES2022} rung for rung: Cepheid distances calibrate
a higher-rung standard candle, which is then applied to Hubble-flow
sources.  Here the standard candle is the population of
individually resolved neutrino-emitting Seyferts rather than
Type~Ia supernovae.  The normalisation of the $L_\nu$--$L_X$
relation plays the role of the supernova absolute magnitude, with
one simplification: in our likelihood it is profiled out
analytically rather than calibrated empirically against an
external population.  The Cepheid rung itself, with its associated
systematics, is retained.

Third, the slope $\beta = 0.67^{+0.16}_{-0.25}$ is a physics result
in its own right.  Corona models in which cosmic-ray injection
traces the coronal dissipation power predict an approximately
linear scaling, $\beta \approx 1$~\cite{Kheirandish2021}; the
contemporary realisation of the classical calorimetric
expectation~\cite{Eichler1979}, and the asymptotic
$pp$-calorimetric limit of the disk-corona
framework~\cite{MKM2020}.  Our data exclude this at ${\sim}2.0\sigma$,
pointing instead to a regime in which neutrino production becomes less efficient per unit $L_X$ as a source brightens.
Several mechanisms could produce such behaviour; Bethe--Heitler
pair losses on disk photons overtaking pion production in the most
luminous coronae~\cite{MKM2020}, source-to-source variation in
cosmic-ray injection efficiency, or sub-dominant non-coronal
contributions to $L_X$ in some sources, and disentangling them
will require a substantially larger sample.  A precise
determination of $\beta$ in the IceCube-Gen2 era would therefore
both sharpen the $H_0$ measurement and provide a population-level
test of the disk-corona neutrino-production framework.

The IceCube-Gen2 forecast in \cref{fig:H0_whisker}
captures the dominant improvement in $H_0$ precision. Complementary inputs will also matter. Additional anchor distances,
from Cepheid, tip-of-the-red-giant-branch, and megamaser surveys, will tighten
the calibrator rung; and access to IceCube's internal likelihood
model with the full 14-year dataset will sharpen $\hat{n}_s$ beyond
what the public release~\cite{IceCube_14yr} allows.\\
\textit{Conclusions.}---We have proposed a neutrino-based method for measuring the Hubble constant from a population of X-ray--bright Seyfert galaxies, and used it to obtain the first determination of $H_0$ with high-energy astrophysical neutrinos, $H_0 = 49^{+40}_{-30}\,\mathrm{km\,s^{-1}\,Mpc^{-1}}$.  The same data deliver the first population-level measurement of the neutrino--X-ray luminosity slope, $\beta = 0.67^{+0.16}_{-0.25}$, disfavoring the calorimetric limit at ${\sim}2.0\sigma$.  Together they establish a new rung of the cosmic distance ladder; one that photons traverse only at the local anchor, immune to the dust, extinction, and radiative-transfer systematics that limit every electromagnetic rung above it.

\begin{acknowledgments}
\textit{Acknoledgements.}--We are grateful to Francis Halzen, Ali Kheirandish and Kohta Murase for useful comments on this manuscript. This work uses public data from the IceCube Neutrino Observatory
(14-year point-source release). The work of G.H is supported by
the Neutrino Theory Network Fellowship with contract
number 726844.
C.A.A. are supported by the Faculty of Arts and Sciences of Harvard University, the National Science Foundation (NSF, CAREER Grant \#2239795), the John Templeton Foundation (Grant \#63651), the Research Corporation for Science Advancement, and the David \& Lucile Packard Foundation.
N.K. is supported by the National Science Foundation (NSF, CAREER Grant \#2239795) and the David \& Lucile Packard Foundation.
\end{acknowledgments}

\clearpage
\onecolumngrid
\begin{center}
{\large \textbf{Supplemental Material:\\[2pt] Measurement of the Hubble constant with high-energy neutrinos}}
\end{center}
\vspace{4pt}
\twocolumngrid
\setcounter{equation}{0}
\setcounter{figure}{0}
\setcounter{table}{0}
\setcounter{section}{0}
\setcounter{secnumdepth}{3}
\renewcommand{\theequation}{S\arabic{equation}}
\renewcommand{\thefigure}{S\arabic{figure}}
\renewcommand{\thetable}{S\Roman{table}}

%

This Supplemental Material accompanies the  Letter and provides a
self-contained description of the analysis.  We begin with the
derivation of the model predictions, including a unified detector
response framework (Sec.~\ref{sec:model}), a correction factor for
redshift-dependent effects on the arriving high-energy neutrino and X-ray fluxes (Sec.~\ref{sec:kcorr}), and the parametrization
equivalence and noise-model justification (Sec.~\ref{sec:noise}).  We
then describe the 12-source catalog and signal-count extraction
(Sec.~\ref{sec:catalog}), the full likelihood and prior specification
(Sec.~\ref{sec:inference}), and the MCMC sampling and convergence
diagnostics (Sec.~\ref{sec:mcmc}).  Section~\ref{sec:skyllh} presents
the per-source profile-likelihood scans and the SkyLLH validation
against published IceCube test statistics.
Sections~\ref{sec:priors} and~\ref{sec:data_info} analyze the impact
of prior choices and identify which parameters are constrained by the
data versus by the priors.  Section~\ref{sec:calibrators} examines the
sensitivity to the choice of Cepheid calibrator;
Sec.~\ref{sec:fisher} the IceCube-Gen2 Fisher forecast; and
Sec.~\ref{sec:tables} the detailed source tables;
Sec.~\ref{sec:pop61} discusses the prospects for extending the
analysis to the broader 61-source IceCube population sample in
the IceCube-Gen2 era.  Finally,
Sec.~\ref{sec:pecvel} details the CMB-frame conversion and the
per-source peculiar-velocity corrections applied to obtain the
primary posterior.

\section{Model derivation and detector response}
\label{sec:model}

The disk-corona model~\cite{MKM2020-SM,Murase2022,Kheirandish2021-SM}
predicts that cosmic rays accelerated by turbulent magnetic fields in AGN
coronae interact with coronal X-ray photons via photomeson ($p\gamma$)
and/or hadronic ($pp$) processes, producing high-energy neutrinos. In the following we describe how this theoretically expected correlation is employed to extract information on $H_{0}$.

\subsection{Luminosity scaling and flux relations}

The fundamental relation linking neutrino flux to X-ray flux at a given source is
\begin{equation}
  L_\nu = \kappa L_X^\beta\,,
  \label{eq:lnu_lx}
\end{equation}
where $\kappa$ is a normalization constant with units of $(\mathrm{erg\,s^{-1}})^{1-\beta}$.

The observed fluxes at Earth are related to luminosities by the inverse-square law:
\begin{equation}
  F_\nu = \frac{L_\nu}{4\pi d^2}\,, \quad
  F_X = \frac{L_X}{4\pi d^2}\,,
\end{equation}
where $d$ is the luminosity distance to the source.

Substituting Eq.~(\ref{eq:lnu_lx}) into the neutrino flux expression:
\begin{equation}
  F_\nu = \frac{\kappa L_X^\beta}{4\pi d^2} = \frac{\kappa (4\pi d^2 F_X)^\beta}{4\pi d^2}
  = \kappa (4\pi)^{\beta-1} d^{2(\beta-1)} F_X^\beta\,.
  \label{eq:flux_scaling}
\end{equation}

This reveals the fundamental distance sensitivity: the expected signal scales as $d^{2(\beta-1)}$.
When $\beta = 1$ exactly, this exponent vanishes and the method is blind to $H_0$.

Equation~(\ref{eq:flux_scaling}) contains the two population-level
unknowns of the method; the normalization $\kappa$ and the distance
$d$, multiplicatively coupled through the slope $\beta$.  For a
Hubble-flow source with $d = c z / H_0$, Eq.~(\ref{eq:flux_scaling})
becomes $F_\nu \propto \kappa\,(c z / H_0)^{2(\beta - 1)}\, F_X^\beta$,
so the transformation
\begin{equation}
  \kappa \;\to\; \lambda\,\kappa\,,\qquad
  H_0 \;\to\; \lambda^{1/[2(\beta-1)]}\,H_0
  \label{eq:kappa_H0_degeneracy}
\end{equation}
leaves every predicted flux unchanged.  Away from the calorimetric limit
$\beta = 1$, $\kappa$ and $H_0$ are therefore perfectly degenerate:
observations of Hubble-flow sources alone cannot determine either.  This
is the direct analogue of the classical distance-ladder problem, in which
the absolute magnitude of a standard candle cannot be measured from
Hubble-flow galaxies without an external, $H_0$-independent anchor.

A non-redshift distance to a single source in the sample breaks this
degeneracy.  For NGC~1068, whose Cepheid+TRGB envelope distance
$d_{1068} = 10.93 \pm 0.74\,\mathrm{Mpc}$ is set externally and does not scale with $H_0$, Eq.~(\ref{eq:flux_scaling}) reads
\begin{equation}
  \kappa \;=\; \frac{F_\nu^{\rm NGC\,1068}}{(4\pi)^{\beta-1}\,
        d_{1068}^{\,2(\beta-1)}\, F_{X,\rm NGC\,1068}^{\,\beta}}\,,
  \label{eq:kappa_from_anchor}
\end{equation}
so that; once $\beta$ is fixed by the shape of the $L_\nu$--$L_X$
relation across the population, -$\kappa$ is analytically determined by
the anchor alone.  With $\kappa$ pinned, the eleven remaining
Hubble-flow sources constrain $H_0$ jointly through their
$d = c z / H_0$ dependence in Eq.~(\ref{eq:flux_scaling}).  The
physical role of the anchor is therefore closely analogous to the LMC in
the classical Cepheid ladder: it supplies the absolute distance scale
that turns a purely relative measurement of the $L_\nu$--$L_X$
correlation into an absolute one, so that $H_0$ can be read off the
Hubble flow.

\subsection{Connection to measured signal counts}

IceCube measures event counts $n_s$, not fluxes directly.  The
predicted signal count from source $i$ is related to the neutrino
energy flux received at Earth by
\begin{equation}
  n_{s,i} = C_i \times F_{\nu,i}\,,
  \label{eq:ns_via_C}
\end{equation}
where $F_{\nu,i}$ (units erg\,cm$^{-2}$\,s$^{-1}$) is the neutrino
energy flux integrated over the IceCube-sensitive band
$[E_{\rm min}, E_{\rm max}]$, and $C_i$ is a source-specific
detector response factor with units of
$(\mathrm{erg\,cm^{-2}\,s^{-1}})^{-1}$.  We define $C_i$
operationally as the expected signal counts per unit integrated
neutrino energy flux from a source at declination $\delta_i$ with
spectral index $\gamma_{\nu,i}$; it is computed for each source from
IceCube's public 14-year effective-area tables and per-season
livetimes.  For a power-law spectrum this reduces to
\begin{equation}
  C_i = \frac{I_A(\gamma_{\nu,i}, \delta_i)}{I_E(\gamma_{\nu,i})}\,,
  \label{eq:Ci_def}
\end{equation}
where
\begin{equation}
  I_A(\gamma_\nu, \delta) = \sum_{\rm seasons} T_s \int_{E_{\rm min}}^{E_{\rm max}}
    A_{\rm eff,s}(E,\delta) \left(\tfrac{E}{E_{\rm ref}}\right)^{-\gamma_\nu} dE,
  \label{eq:IA}
\end{equation}
is the effective area integrated over energy (weighted by the assumed
spectral shape) and
\begin{equation}
  I_E(\gamma_\nu) = \int_{E_{\rm min}}^{E_{\rm max}} E \left(\tfrac{E}{E_{\rm ref}}\right)^{-\gamma_\nu} dE,
  \label{eq:IE}
\end{equation}
is the corresponding energy-flux normalisation integral. 
The number of signal events, $n_s = C_i\,F_\nu$, is manifestly dimensionless when $F_\nu$ is
the band-integrated energy flux in $\mathrm{GeV}\,\mathrm{cm}^{-2}\,\mathrm{s}^{-1}$.
This
decomposition is a book-keeping convenience: only the numerical values
of $C_i$ (declination-dependent through $A_{\rm eff}$, spectral-index
dependent through both integrals) enter the likelihood.

Combining Eq.~(\ref{eq:ns_via_C}) with the $L_\nu$--$L_X$
correlation, taking $L_\nu = \kappa L_X^{\,\beta}$ to mean the
band-integrated luminosity in the IceCube-sensitive range, gives
\begin{equation}
  F_{\nu,i} = \frac{L_{\nu,i}}{4\pi d_i^2} = \frac{\kappa\, L_{X,i}^{\,\beta}}{4\pi d_i^2}\,,
  \label{eq:Fnu_from_LX}
\end{equation}
with $\kappa$ in units of $(\mathrm{erg\,s^{-1}})^{1-\beta}$.  Using
$L_X = 4\pi d^2 F_X$ we can express this in terms of the observed
X-ray flux:
\begin{equation}
  F_{\nu,i} = \kappa\,(4\pi)^{\beta-1}\, d_i^{2(\beta-1)}\, F_{X,i}^{\,\beta}\,.
  \label{eq:Fnu_from_FX}
\end{equation}
Substituting into Eq.~(\ref{eq:ns_via_C}), the predicted signal count
for source $i$ as a function of the model parameters is
\begin{equation}
  \mu_i(H_0, \beta) = A \cdot C_i(\gamma_{\nu,i}, \delta_i) \cdot F_{X,i}^{\,\beta} \cdot d_{L,i}^{\,2(\beta-1)}(H_0)\,,
  \label{eq:mu_unified}
\end{equation}
where $A \equiv \kappa\,(4\pi)^{\beta-1}$ absorbs the population-level
normalisation and the geometric factor, and $d_{L,i}(H_0)$ is the
$\Lambda$CDM luminosity distance for Hubble-flow sources or the
externally-set geometric distance for the anchor.  The detector
response $C_i$ is redshift-independent and factors cleanly from the
distance dependence, which is what allows the method to constrain
$H_0$: variation in $\mu_i$ across the sample at fixed $A$ and
$\beta$ is driven by the $d_{L,i}^{\,2(\beta-1)}$ lever arm, and
this is where the $H_0$ information enters.

\subsection{Luminosity distance calculation}

Luminosity distances are computed in flat $\Lambda$CDM with
$\Omega_m = 0.315$ and $\Omega_\Lambda = 0.685$ via numerical
integration of
\begin{equation}
  d_L(z, H_0) = (1+z) \int_0^z \frac{c\,dz'}{H_0\,
    \sqrt{\Omega_m(1+z')^3 + \Omega_\Lambda}}\,.
  \label{eq:dL}
\end{equation}

\section{Theoretical expectations for the slope $\beta$}
\label{sec:beta_theory}
As revealed in \cref{eq:mu_unified}, the value of $\beta$ plays a central role in the ability to constrain $H_0$.
Specifically, $\beta\neq 1$ is required to derive any information on $H_0$.
This section explores the different astrophysical environments that can lead to deviations from the calorimetric limit, i.e. $\beta=1$.
The regime $\beta>1$ can occur for $p\gamma$ collisions in the thin-target regime~\cite{Karavola2025}.
The regime $\beta<1$ is less explored in the literature, though it is the regime preferred by the fit in the Letter.
We therefore discuss two mechanisms, Bethe-Heitler suppression and synchrotron cooling of pions and muons, that could plausibly lead to $\beta<1$.

\subsection{The calorimetric limit $\beta = 1$}
The calorimetric limit refers to the regime in which the AGN corona is optically thick to $p\gamma$ interactions.
Here, the neutrino luminosity is expected to scale with the X-ray luminosity as~\cite{MuraseGuettaAhlers2016,Karavola2025}
\begin{equation} \label{eq:Lnu_LX}
    E_\nu \mathcal{L}_\nu \approx \frac{3K}{4(1+K)}f_\pi E_p \mathcal{L}_p,
\end{equation}
where $K$ is the average ratio of charged to neutral pions with $K=1 (2)$ for $p\gamma$ ($pp$) interactions, $f_\pi$ is the pion production efficiency, $\mathcal{L}_p$ is the cosmic ray luminosity of the source, and $E_\nu~(E_p)$ is the neutrino (cosmic ray) energy.
The calorimetric limit explicitly refers to $f_\pi \to 1$, i.e. all injected protons convert into pions.
In the typical case that $L_p \propto L_X$, which is true when the cosmic ray loading factor is constant~\cite{MKM2020-SM,MuraseStecker2023}, we recover a linear dependence $L_\nu \propto L_X$.

\subsection{The optically-thin regime $\beta > 1$}

In the optically-thin regime, pion production via $p\gamma$ is not efficient in the coronal region.
It is instead given by the expression~\cite{Karavola2025}
\begin{equation}
    f_\pi = t_{p\gamma}^{-1} R/c,
\end{equation}
where $R$ is the coronal radius and $t_{p\gamma}$ is the timescale of $p\gamma$ interactions.
Because protons are scattering off of the same photons that contribute to $L_X$, this term can be written as~\cite{Karavola2025}
\begin{equation}
  t_{p\gamma}^{-1} R/c \approx 0.3 \frac{L_X}{L_{\rm Edd}} \frac{R_S}{R} \frac{\min(E_p,E_p^*)}{25\,{\rm TeV}},
\end{equation}
where $L_{\rm Edd}$ is the Eddington luminosity, $R_S$ is the Schwartschild radius of the central black hole, $E_p$ is the proton energy, and $E_p^*$ is the threshold proton energy for $p\gamma$ interactions with the lowest energy coronal X-rays.
Plugging this back into \cref{eq:Lnu_LX}, we find a predicted relationship $L_\nu \propto L_X^2 / L_{\rm Edd}$, corresponding to $\beta=2$.
This expression is valid below a critical Eddington ratio $L_X/L_{\rm Edd} \sim 10^{-2}-10^{-1}$~\cite{Karavola2025}.

\subsection{The $\beta < 1$ regime}

We now turn to sub-linear scaling $\beta<1$, the regime currently preferred by IceCube data.
Theoretical predictions for this regime rely on the fact that IceCube measures $L_\nu$ in a specific band for each source, e.g. $0.1\,{\rm TeV} \lesssim E_\nu \lesssim 10\,{\rm TeV}$ for NGC~1068.
If an energy loss process pushes the neutrino flux out of this band and becomes more efficient with $L_X$, it could lead to sublinear scaling of $L_\nu$ with $L_X$.
There are at least two such energy loss mechanisms: Bethe-Heitler suppression and synchrotron cooling of pions and muons.

Bethe-Heitler (BH) suppression refers to proton pair production on disk (UV) and coronal (X-ray) photons, $p\gamma \to p e^+ e^-$.
The threshold for this processes is much lower than that of photomeson production~\cite{MuraseStecker2023}.
Therefore, BH losses can happen on disk photons while photomeson production may be restricted to $p\gamma$ and/or $pp$ in the corona.
Depending on the energy spectrum of disk and coronal photons, BH losses can become stronger with increasing $L_X$, pushing the neutrino flux out of the IceCube sensitive range.
This is exactly the effect expected in turbulent AGN corona, which were found to predict $\beta \approx 0.8$ in \cite{Fiorillo2025}.

Significant synchrotron cooling of pions and muons before decaying can also lead to suppressed neutrino fluxes.
This is because such cooling impacts $\pi^\pm$ and $\mu^\pm$, the generators of the neutrino luminosity, without impacting $\pi^0$, the generator of the photon luminosity.
Since synchrotron energy losses increase with $B^2$, AGN with higher $L_X$ (and thus higher magnetic fields) will feature stronger synchrotron cooling.
For this mechanism to be efficient, the synchrotron cooling timescale must be shorter than the decay timescale of charged pions and muons, requiring coronal magnetic fields of the order $B \sim 10^7\,{\rm G}$~\cite{Blanco2025}.

\section{$k$-correction for redshift effects}
\label{sec:kcorr}

For sources at non-negligible redshift, the observed flux differs from the rest-frame flux
due to the relativistic $(1+z)$ Doppler and cosmological redshift effects. The primary
three-dimensional analysis ($\{H_0,\beta,d_{1068}\}$) fixes the $k$-correction factor $K_i \equiv 1$;
this section derives the exact $k$-correction and quantifies the residual systematic
uncertainty via a dedicated sensitivity run in which $\gamma_X$ is sampled as a free
parameter and $K_i$ is applied source-by-source.

\subsection{Rest-frame vs.\ observer-frame fluxes}
For a source at redshift $z$ with a power-law neutrino spectrum
$dL_\nu/dE \propto E^{-\gamma_\nu}$, the band-integrated energy
fluxes in matched rest-frame and observer-frame bands are related by
the standard $k$-correction
\begin{equation}
  F_\nu^{\rm rest} = (1+z)^{\gamma_\nu - 2}\,F_\nu^{\rm obs}\,.
  \label{eq:kcorr_nu}
\end{equation}
The exponent decomposes as the sum of three effects:
a factor $(1+z)^{\gamma_\nu}$ from evaluating the power law in the
shifted band, $(1+z)^{-1}$ from bandwidth compression
($dE_{\rm rest} = (1+z)\,dE_{\rm obs}$), and a further $(1+z)^{-1}$
from the per-neutrino energy redshift between emission and detection.
The first two combine into $(1+z)^{\gamma_\nu - 1}$, the correction
for the band-integrated number flux; adding the third gives
$(1+z)^{\gamma_\nu - 2}$ for the energy flux, as above.

\subsection{Application to the neutrino-X-ray relation}

The disk-corona relation in the rest frame of the source is:
\begin{equation}
  L_{\nu,{\rm rest}} = \kappa L_{X,{\rm rest}}^\beta\,.
\end{equation}

Applying the $k$-correction to both bands, substituting into the rest-frame relation, and simplifying, yields:
\begin{equation}
  F_{\nu,{\rm obs}} = \kappa (4\pi d_L^2)^{\beta-1} (1+z)^{\beta(\gamma_X-2) - (\gamma_\nu-2)} F_{X,{\rm obs}}^\beta\,.
  \label{eq:fnu_kcorr}
\end{equation}

The $k$-correction introduces an additional factor:
\begin{equation}
  K = (1+z)^{\beta(\gamma_X-2) - (\gamma_\nu-2)} = (1+z)^{\beta\gamma_X - 2\beta - \gamma_\nu + 2}\,.
  \label{eq:kcorr_factor}
\end{equation}

For the baseline parameters $\gamma_X \approx 1.8$, $\gamma_\nu \approx 3.0$, and $\beta \approx 0.70$:
\begin{align}
  K= (1+z)^{-1.14}\,.
\end{align}

At $z = 0.05$ (near the upper end of our sample's redshift range), the correction is of $\sim 5\%$. At $z = 0.10$, the correction is of $\sim 10\%$.

\subsection{Impact on $H_0$ inference}

The primary analysis fixes $K_i \equiv 1$ and samples the three-dimensional space
$\{H_0,\beta,d_{1068}\}$.  To verify that this approximation is safe, we run a dedicated
four-dimensional sensitivity chain in which $\gamma_X$ is added as a free parameter with
flat prior $\gamma_X \in [1, 3]$ and the source-dependent $K_i(z_i;\beta,\gamma_X,\gamma_{\nu,i})$
of Eq.~(\ref{eq:kcorr_factor}) is applied at every likelihood evaluation.  The neutrino spectral
index $\gamma_{\nu,i}$ for each source enters through the SkyLLH detector response factor
$C_i$ (Sec.~\ref{sec:skyllh}) at its best-fit value $\hat{\gamma}_{\nu,i}$, with the
associated uncertainty propagated into $\sigma(\hat{n}_s)_i$.

The sensitivity chain returns $\gamma_X = 2.0 \pm 0.7$, fully consistent with the flat
prior and with typical BAT-selected Seyfert spectra.  For this posterior the $k$-correction
factor $K_i$ deviates from unity by less than $3\%$ for the great majority of sources in
the $z < 0.06$ sample.  Comparing the $K_i\!=\!1$ primary chain to the $K_i\!\neq\!1$
sensitivity chain shifts the $H_0$ posterior median by less than $2\,\mathrm{km\,s^{-1}\,Mpc^{-1}}$
and the slope $\beta$ by less than $0.01$, in both cases well below the statistical
uncertainty.

\section{Noise model and parametrization equivalence}
\label{sec:noise}

The IceCube analyses quote best-fit signal counts $\hat{n}_s$ obtained from profile likelihoods
over the spectral index $\gamma_\nu$. The profile likelihood produces an uncertainty $\sigma(\hat{n}_s)$
derived from the Wilks' theorem contour (the $1\sigma$ contour satisfies $\Delta(-2\ln\mathcal{L}) = 1$). In the following we discuss how this information is translated into our likelihood model.

\subsection{Parametrization: counts vs.\ luminosity}

Since the predicted neutrino luminosity relates to the signal count by Eq.~(\ref{eq:mu_unified}):
\begin{equation}
  L_\nu = \frac{4\pi d^2  \, n_s}{C_i}\,,
\end{equation}
where $4\pi d^2$, and $C_i$ are constants for a
given source and spectral index $\gamma_\nu$.

Therefore, if the noise in $n_s$ is Gaussian with standard deviation $\sigma(n_s)$,
the noise in $L_\nu$ is also Gaussian with standard deviation:
\begin{equation}
  \sigma(L_\nu) = \frac{4\pi d^2}{C_i} \, \sigma(n_s)\,.
\end{equation}

The likelihood in $L_\nu$ space, with correct error propagation, is:
\begin{equation}
  \ln\mathcal{L}(L_\nu^{\rm obs} | L_\nu^{\rm pred}) = -\frac{1}{2} \frac{(L_\nu^{\rm obs} - L_\nu^{\rm pred})^2}{\sigma(L_\nu)^2}\,.
\end{equation}

Substituting $L_\nu^{\rm obs} = \alpha_i n_s^{\rm obs}$, $L_\nu^{\rm pred} = \alpha_i n_s^{\rm pred}$,
and $\sigma(L_\nu) = \alpha_i \sigma(n_s)$:
\begin{equation}
  \ln\mathcal{L} = -\frac{1}{2} \frac{\alpha_i^2 (n_s^{\rm obs} - n_s^{\rm pred})^2}{\alpha_i^2 \sigma(n_s)^2}
  = -\frac{1}{2} \frac{(n_s^{\rm obs} - n_s^{\rm pred})^2}{\sigma(n_s)^2}\,.
\end{equation}

The $\alpha_i$ factors (which depend on $d_i$, $C_i$, and constants) cancel exactly.
Therefore, parametrizing in terms of signal counts $n_s$ with Gaussian likelihood is
equivalent to parametrizing in terms of luminosity $L_\nu$ with error propagation.
Because $\alpha_i$ depends on the cosmological parameters, the
normalized luminosity-space density differs from the count-space
density by the change-of-variables factor
$|\partial L_\nu/\partial n_s| = \alpha_i$, which cancels the
$\alpha_i$ in $\sigma(L_\nu)$ and leaves the parameter-free
count-space normalization $-\ln\sigma(n_s)$, as used in
Eq.~(\ref{eq:Li_full}); all likelihood evaluations are performed
with respect to the count-space measure.

\section{Source catalog and signal-count extraction}
\label{sec:catalog}

Our primary analysis uses 12 X-ray-bright Seyfert galaxies, three of
serves as the primary non-redshift-distance anchor (NGC~1068 Cepheid+TRGB),
while the remaining eleven lie in the Hubble flow.  This section describes how the sample
was selected, how signal counts were extracted from the IceCube
14-year public point-source data, and how the Cepheid distances were
adopted.

\subsection{Source selection}

The starting catalog is the 47 X-ray-bright Seyfert galaxies in the
IceCube 13-year population analysis~\cite{IceCube_Xray_2025-SM}, which
were drawn from the BAT AGN Spectroscopic Survey
(BASS)~\cite{BASS-SM,SwiftBAT-SM} on the basis of their \emph{intrinsic}
(absorption-corrected) 20--50\,keV X-ray flux, with the harder band
deliberately chosen by IceCube to minimise the absorption correction
for Compton-thick sources.  All $F_X$ values used in our analysis
are taken verbatim from the $F^{\rm intr}_{20\text{--}50\,\rm keV}$
column of Table~D.4 of Ref.~\cite{IceCube_Xray_2025-SM}; that column is
derived from the BASS DR2 catalog~\cite{BASS-SM}.  For NGC~1068, NGC~1194, and NGC~3079
(Compton-thick, $N_H \gtrsim 10^{24}\,\mathrm{cm^{-2}}$) the de-absorption
correction raises $F_X$ above the directly observed 14--195\,keV BAT
flux by factors of $\sim 2$--$4$; we capture the residual
de-absorption uncertainty by inflating $\sigma_{F_X}/F_X$ to
$20$--$25\%$ for these three sources (vs.\ $15\%$ for Compton-thin
sources).  Sources were
observed at $\delta > -5^\circ$ with the through-going muon-track
event selection.  From this parent sample we retain the twelve
sources with the highest individual significances in the IceCube
13-year analysis~\cite{IceCube_Xray_2025-SM}, among them the three
non-redshift-distance anchors NGC~1068, NGC~4151, and NGC~7469; the
first two are also the most significant sources in our own fits.  Because our signal
counts are re-extracted from the 14-year public data
(Sec.~\ref{sec:skyllh}), the test statistics in
Table~\ref{tab:sources_full} differ from the 13-year significances
that define the selection.  The resulting catalog is listed in
Table~I of the main Letter and reproduced with extended
detector-response columns in Table~\ref{tab:sources_full} below.

We do not attempt to incorporate the southern-sky ESTES sample
analyzed in Ref.~\cite{IceCube_South_2026} into the primary 12-source
catalog: those sources are reported with disk-corona model
predictions rather than with per-source profile likelihoods over
$(n_s, \gamma_\nu)$, which prevents a uniform statistical treatment
within our likelihood framework (Sec.~\ref{sec:inference}).

\subsection{SkyLLH signal-count extraction}

Rather than adopting the population-stacked $\hat{n}_s$ values
quoted in Ref.~\cite{IceCube_Xray_2025-SM}, we re-extract per-source
signal counts directly from the IceCube 14-year public point-source
data~\cite{IceCube_14yr-SM} using the SkyLLH unbinned likelihood
framework~\cite{SkyLLH-SM}.  This step is essential because the
inference in Sec.~\ref{sec:inference} requires not only $\hat{n}_s$
but also the full uncertainty $\sigma(\hat{n}_s)_i$ from the
profile likelihood and the best-fit per-source spectral index
$\hat{\gamma}_{\nu,i}$, neither of which is directly tabulated in the
population paper.

For each source we evaluate the SkyLLH unbinned point-source
likelihood on a uniform grid of $(n_s, \gamma_\nu)$ with
$n_s \in [0, n_{s,\max}]$ (where $n_{s,\max}$ is set
source-by-source to safely encompass the $3\sigma$ contour) and
$\gamma_\nu \in [1.0, 4.0]$ at resolution $60 \times 50$.  The
best-fit point in this two-dimensional likelihood surface defines
$\hat{n}_s$ and $\hat{\gamma}_\nu$, and the test statistic is
\begin{equation}
  {\rm TS}_i = -2 \left[\ln\mathcal{L}_i(0, \gamma_\nu^*) -
                       \ln\mathcal{L}_i(\hat{n}_s,
                       \hat{\gamma}_\nu)\right],
\end{equation}
where $\gamma_\nu^*$ is the value that maximizes the null hypothesis.
The one-dimensional profile likelihood
$\Delta(-2\ln\mathcal{L})(n_s)$, obtained by minimizing over
$\gamma_\nu$ at each $n_s$, defines the $1\sigma$ uncertainty
$\sigma(\hat{n}_s)_i$ from the $\Delta(-2\ln\mathcal{L}) = 1$
contour.  Section~\ref{sec:skyllh} shows the full set of profile
scans and validates the SkyLLH implementation against the published
IceCube 14-year point-source TS catalog.

\subsection{Non-redshift distance anchor}\label{sec:anchors}

The primary analysis uses a single non-redshift distance anchor:
NGC~1068.  This choice is motivated by four complementary considerations.
\emph{(i) Statistical.}  NGC~1068 is the only source in our sample with an
individually significant neutrino excess at the ${\sim}4.2\sigma$
level~\cite{IceCube_NGC1068_2022-SM}; the remaining eleven Seyferts contribute
as an ensemble at ${\sim}3.3\sigma$~\cite{IceCube_Xray_2025-SM}, with per-source
test statistics $\mathrm{TS}\lesssim 12$ (Table~\ref{tab:sources_full}).
Any anchor tension therefore couples most strongly to the source whose
neutrino luminosity is best determined; using it as the calibrator minimises
the propagated uncertainty on the $L_\nu$--$L_X$ normalisation.
\emph{(ii) Proximity---the LMC of the sample.}  At
$d_{1068} = 10.93\,\mathrm{Mpc}$, NGC~1068 is the closest galaxy in our
sample.  Just as the Large Magellanic Cloud plays this role in the
SH0ES Cepheid$\to$SN~Ia ladder, proximity is what makes
precise geometric distance measurement possible in the first place: it is
the only Seyfert in the sample in which individual Cepheid variables and
tip-of-the-red-giant-branch stars are simultaneously resolvable, giving
\emph{two} independent geometric distance measurements---Cepheid variables
from~\cite{NGC1068_Cepheid-SM} and TRGB photometry---which we combine into the
envelope $d_{1068} = 10.93 \pm 0.74\,\mathrm{Mpc}$ (6.8\% fractional
uncertainty).  This is smaller than the systematic-limited uncertainty of every
other geometric distance in the sample: NGC~4151 (17.4~Mpc): 17\% Cepheid+dust
envelope; NGC~7469 (58.2~Mpc): single Type~Ia calibration with ${\sim}3\text{--}5\%$
ladder systematics on top of the geometric error; NGC~3079 (16.7~Mpc): TRGB
alone, ${\sim}8\%$~\cite{Kuo2011}.  Anchoring at the closest source with the
best-measured distance is standard practice in every rung of the classical
distance ladder and is what makes the calibration systematically defensible.
\emph{(iii) Physical.}  NGC~1068 is the archetype obscured Seyfert-2 in which
the disk-corona neutrino production model was originally developed and
successfully confronted with data.
Anchoring the $L_\nu$--$L_X$ correlation at the source whose X-ray corona is
best constrained observationally (Compton-thick column density, coronal
temperature, electron optical depth all measured in high-quality X-ray
spectra) ties the calibration to the source where the model is most tested.
\emph{(iv) Independence.}  The Cepheid and TRGB distance ladders are calibrated
against geometric anchors---LMC eclipsing binaries, water masers---that are
fully independent of the local supernova-based distance scale.  This makes
the resulting $H_0$ measurement genuinely independent of both the SH0ES
Cepheid$\to$SN~Ia ladder and the CMB, avoiding cross-contamination of
systematics.
The remaining eleven Seyferts are treated as Hubble-flow
sources, with distances $d_i = c z_i / H_0$ set by their redshifts and
the sampled $H_0$.  Sensitivity to adding secondary anchors is
explored in Sec.~\ref{sec:calibrators}.

\paragraph{NGC~1068 (Cepheid + tip-of-the-red-giant-branch envelope).}
We adopt an envelope prior that combines the Cepheid distance modulus of Ref. ~\cite{NGC1068_Cepheid-SM} with the tip-of-the-red-giant-branch
distance to the same galaxy, yielding an
envelope $d_{1068} = 10.93 \pm 0.74\,\mathrm{Mpc}$ that
accommodates both the Cepheid distance $10.72 \pm 0.52\,\mathrm{Mpc}$
\cite{NGC1068_Cepheid-SM} and the independent TRGB distance
$11.14 \pm 0.54\,\mathrm{Mpc}$~\cite{NGC1068_TRGB-SM} at the $1\sigma$ level.  The envelope width is broader than either error bar individually because it absorbs the systematic offset between the two independent standard candles.  This anchor enters
the MCMC as the sampled parameter $d_{1068}$, constrained by a truncated
Gaussian prior with the above width; its uncertainty propagates fully
into the $H_0$ posterior.

\paragraph{Sources considered as secondary anchors (sensitivity only).}
Two additional sources in our sample have published non-redshift
distances that we treat as sensitivity tests rather than primary
anchors:
\emph{(i)} NGC~4151 has a Cepheid+dust-parallax envelope distance
$d_{4151} = 17.4 \pm 3.0\,\mathrm{Mpc}$~\cite{NGC4151_Cepheid};
\emph{(ii)} NGC~7469 has a Type~Ia supernova distance from SN~2008ec,
$d_{7469} = 58.2 \pm 1.9\,\mathrm{Mpc}$~\cite{Koshida2017}. Sec.~\ref{sec:calibrators}
quantifies how the $H_0$ posterior shifts when NGC~4151 and NGC~7469
are added as secondary anchors.

\section{Statistical inference}
\label{sec:inference}

\subsection{Likelihood}

For each of the 12 sources in the catalog, the model prediction for
the neutrino luminosity follows from the unified detector response
framework derived in Sec.~\ref{sec:model}:
\begin{equation}
  L_{\nu,i}^{\rm pred}(H_0, \beta, d_{1068})
  = \kappa\, L_{X,i}^{\,\beta}\, K_i(z_i; \gamma_X, \hat\gamma_{\nu,i})\,,
  \label{eq:Lnu_pred_full}
\end{equation}
where $L_{X,i} = 4\pi\, d_{L,i}^2\, F_{X,i}$ uses the appropriate
distance ($d_{1068}$ for NGC~1068 and $d_L(z_i, H_0)$
for the other eleven Hubble-flow sources).  In the primary analysis we fix $K_i \equiv 1$;
the dedicated sensitivity run that frees $\gamma_X$ and applies the source-dependent
$k$-correction $K_i(z_i;\beta,\gamma_X,\gamma_{\nu,i})$ is documented in
Sec.~\ref{sec:kcorr}.

The corresponding inferred neutrino luminosity from the data is
\begin{equation}
  \hat{L}_{\nu,i} = \frac{4\pi d_{L,i}^2}{C_i(\hat\gamma_{\nu,i},\delta_i)}\,
                    \hat{n}_{s,i}\,,
\end{equation}
with $C_i$ the per-source detector response factor evaluated at the
best-fit spectral index from the SkyLLH scan.  The propagated
luminosity uncertainty is
\begin{equation}
  \sigma_{L,i}^2 = \left(\frac{4\pi d_{L,i}^2}{C_i}\right)^2
                    \sigma(\hat{n}_s)_i^2
                    + (\sigma_{F_X,i}\, L_{\nu,i}^{\rm pred})^2\,,
  \label{eq:sigma_L_full}
\end{equation}
where the first term is the propagated SkyLLH uncertainty and the
second term is the contribution from the fractional X-ray flux
uncertainty $\sigma_{F_X,i}/F_{X,i}$, held at $25\%$ uniformly for
all sources (Table~I of the main Letter).  The likelihood for
source $i$ is Gaussian:
\begin{equation}
  \ln\mathcal{L}_i = -\frac{1}{2}\,
    \frac{\big(\hat L_{\nu,i} - L_{\nu,i}^{\rm pred}\big)^2}
         {\sigma_{L,i}^2}
    -\ln\sigma_{n,i}\,,
  \label{eq:Li_full}
\end{equation}
where $\sigma_{n,i} \equiv C_i\,\sigma_{L,i}/(4\pi d_{L,i}^{2})$ is
the count-space width, and the total log-likelihood is the sum over
all 12 sources.  The likelihood is defined and evaluated in count
space, whose statistical width $\sigma(\hat{n}_s)_i$ is
parameter-independent, the conversion between counts and
luminosities introducing no additional distance dependence: as shown
in Sec.~\ref{sec:noise}, the residual is identical in the two
parametrizations, and the Jacobian of the parameter-dependent
counts-to-luminosity transformation cancels the distance factors in
the Gaussian normalization.

The second term in Eq.~(\ref{eq:sigma_L_full}) already plays the
role of an intrinsic-scatter term: a fractional, multiplicative-on-%
prediction Gaussian variance.  An intrinsic scatter $\sigma_{\rm int}$
in the $L_\nu$--$L_X$ relation would enter the likelihood in exactly
the same form, so the two are perfectly degenerate and cannot be
separated by the data; a $25\%$ flux uncertainty is equivalent to
$0.25/\ln 10 \approx 0.11$~dex of scatter.  We therefore do not fit a
separate $\sigma_{\rm int}$ parameter; the single $25\%$ X-ray flux
uncertainty absorbs both the measurement error and any intrinsic
astrophysical scatter.  With only 12 sources a free scatter parameter
would in any case be unconstrained and largely degenerate with the
slope $\beta$.

\subsection{Profiling of the normalization}

The overall neutrino normalization $\kappa$ enters
Eq.~(\ref{eq:Lnu_pred_full}) only as a multiplicative constant.
Setting
\begin{equation}
  \frac{\partial \ln\mathcal{L}}{\partial \kappa} = 0
\end{equation}
yields, at fixed variance, the weighted-least-squares value
\begin{equation}
  \kappa^* = \frac{\sum_i L_{X,i}^\beta \hat{L}_{\nu,i}/
                       \sigma_{L,i}^2}
                  {\sum_i (L_{X,i}^\beta)^2/
                       \sigma_{L,i}^2}\,,
  \label{eq:kappa_profile}
\end{equation}
which depends on $\{H_0, \beta, d_{1068}\}$ only through $L_{X,i}$ (with
$K_i \equiv 1$ in the primary analysis).  Since $\sigma_{L,i}$ itself depends on $\kappa$ through the
multiplicative flux term in Eq.~(\ref{eq:sigma_L_full}),
Eq.~(\ref{eq:kappa_profile}) is solved by iterated
reweighting---the variance is re-evaluated at the updated
prediction and Eq.~(\ref{eq:kappa_profile}) reapplied---which
converges within a few iterations.  Substituting the converged
$\kappa^*$ back into Eq.~(\ref{eq:Li_full}) eliminates $\kappa$ from the parameter space
analytically, reducing the sampling problem to three dimensions.
This deterministic profiling ($\kappa$ is maximized over rather
than marginalized) removes the
absolute-normalization degree of freedom that plays no role in the
$H_0$ inference: $\kappa$ is the analog of the supernova absolute
magnitude $M_B$ in the Cepheid--supernova ladder~\cite{Riess2022},
and like $M_B$ it cancels once the ladder is externally anchored.

\subsection{Prior specification}
\label{sec:prior_spec}

The three sampled parameters and their priors are:

\paragraph{$H_0 \in [1, 150]\,\mathrm{km\,s^{-1}\,Mpc^{-1}}$ (uniform).}
Deliberately broad and agnostic, this prior comfortably encompasses
all plausible $H_0$ values from well below the Planck value to
substantially above SH0ES.  Section~\ref{sec:priors} demonstrates
that the posterior is genuinely informed by the data rather than by
the prior boundaries.

\paragraph{$\beta \in [0.1, 3]$ (uniform).}
The slope is allowed to vary over a broad range that encompasses
calorimetric ($\beta = 1$), thin-target $p\gamma$ ($\beta \approx
1.5$--$2$), and saturated/sub-linear regimes ($\beta < 1$). 

\paragraph{$d_{1068} \sim \mathcal{N}(10.93, 0.74^2)\,\mathrm{Mpc}$, truncated to $[8, 14]$.}
NGC~1068 Cepheid+TRGB envelope prior~\cite{NGC1068_Cepheid-SM}; the
uncertainty is broadened beyond the HST Cepheid statistical error to
cover the independent tip-of-the-red-giant-branch determination
$d = 11.14 \pm 0.54\,\mathrm{Mpc}$~\cite{NGC1068_TRGB-SM} of the
NGC~1068 group (see Sec.~\ref{sec:anchors}).

\paragraph{$d_{4151} \sim \mathcal{N}(17.4, 3.0^2)\,\mathrm{Mpc}$, truncated to $[10, 25]$.}
NGC~4151 Cepheid+dust-parallax envelope prior~\cite{NGC4151_Cepheid}; the
width is set so the prior covers both the Cepheid/SN~Ia value
$15.8\pm0.4\,\mathrm{Mpc}$ and the infrared dust-reverberation value
$19.0\pm2.3\,\mathrm{Mpc}$ at $\sim 1\sigma$.

%
%

\section{MCMC sampling and diagnostics}
\label{sec:mcmc}

We sample the joint posterior in the three-dimensional space
$\{H_0, \beta, d_{1068}\}$, where $d_{1068}$ is the distance to single non-redshift-distance anchor NGC~1068, using component-wise
Metropolis--Hastings MCMC with 10 independent walkers initialized
in random points within the prior support.  At each step, each
parameter is proposed independently from a Gaussian centered on the
current value, and accepted or rejected according to the Metropolis
criterion.  This component-wise approach avoids the need to tune a
full covariance matrix.  (The two-anchor configuration that adds
$d_{4151}$ for NGC~4151 is examined in Sec.~\ref{sec:calibrators}.)

Each walker is run for $10^5$ steps with a burn-in of
$3\times 10^4$ steps, leaving $7\times 10^4$ post-burn-in samples
per walker and $7\times 10^5$ total samples used for posterior
inference.  Initial proposal widths are
$[15.0,\ 0.18,\ 0.30]$ for
$\{H_0, \beta, d_{1068}\}$, in their respective units.
During the first half of the burn-in we adapt the proposal widths:
each component's width is multiplied by 1.5 if its acceptance rate
exceeds 0.40 and by 0.5 if it falls below 0.15.  After adaptation
the average per-component acceptance rate is in the range
$0.30$--$0.42$.

\subsection{Convergence diagnostics}
All 10 walkers pass standard convergence checks: the Gelman--Rubin
potential scale reduction factor $\hat R$ satisfies $\hat R \in [1.00,\ 1.03]$
for each parameter, well below the conventional $\hat R < 1.10$ threshold, and
the integrated autocorrelation times ($\tau_{H_0}\!\approx\!280$,
$\tau_\beta\!\approx\!65$, $\tau_{d_1}\!\approx\!30$ steps) give
$\gtrsim 4\times 10^3$ effective independent samples in the combined chain,
more than enough for the reported percentile summaries.

\subsection{Joint posteriors}

\begin{figure}[t]
  \centering
  \includegraphics[width=\columnwidth]{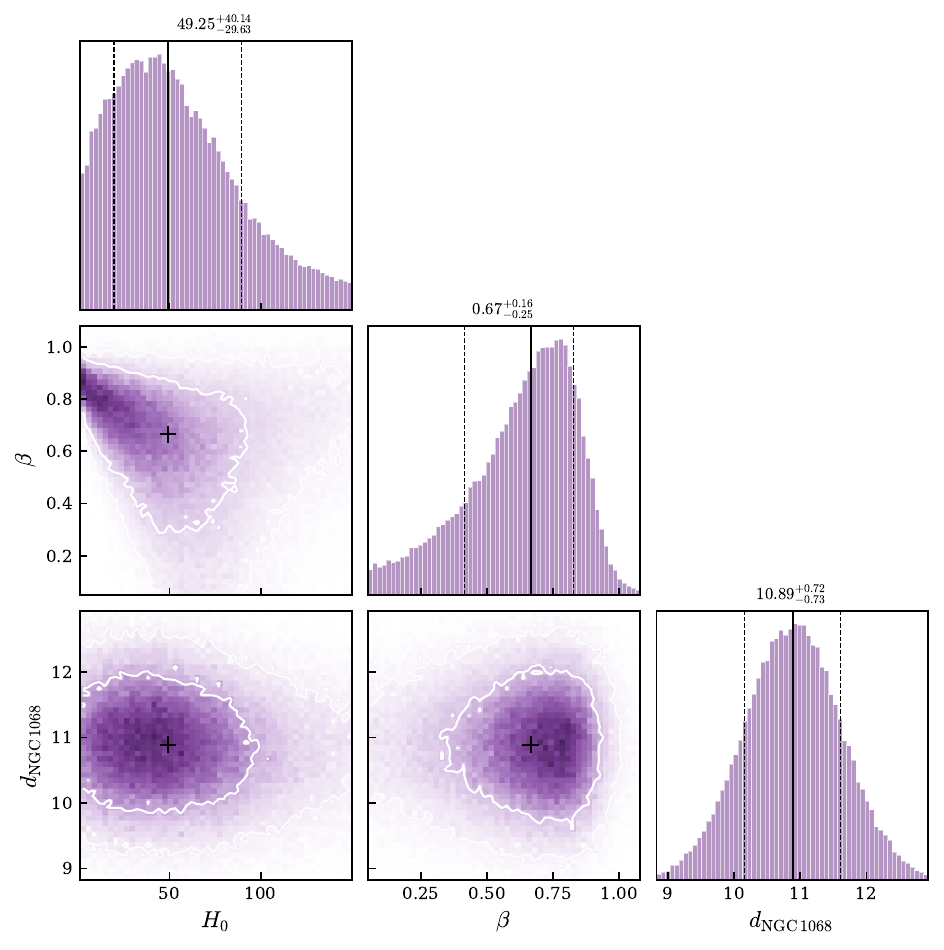}
  \caption{Joint posterior in the $(H_0, \beta)$ plane (combined
    walker output, primary 1-anchor analysis).  Contours mark
    $68\%$ and $95\%$ credible
    regions.  Vertical dashed lines indicate the Planck (blue,
    $67.4\,\mathrm{km\,s^{-1}\,Mpc^{-1}}$) and SH0ES (red,
    $73.0\,\mathrm{km\,s^{-1}\,Mpc^{-1}}$) values; the horizontal dashed
    line marks the calorimetric value $\beta = 1$, which is excluded
    by the data at ${\sim}2\sigma$.}
  \label{fig:corner}
\end{figure}

The full set of marginal and joint posteriors is shown in
Fig.~\ref{fig:corner}.  The most important correlation is the
negative degeneracy between $H_0$ and $\beta$ (correlation
coefficient $\approx -0.5$), reflecting the partial degeneracy
between distance and slope at fixed observed flux: a higher $H_0$
(shorter Hubble-flow distances) can be partly compensated by a
smaller deviation of $\beta$ from unity.  In the dedicated sensitivity
chain of Sec.~\ref{sec:gammaX_sens}, $\gamma_X$ is essentially uncorrelated
with $H_0$ ($|\rho| < 0.1$), confirming that the $k$-correction does not
significantly bias the $H_0$ inference and supporting the $K_i \equiv 1$
choice in the primary three-dimensional chain.  The non-redshift anchor distance $d_{1068}$
shows only a
mild correlation with $H_0$ (the absolute scale that it sets is
shared with $\kappa$, which is profiled out analytically).

\section{Profile likelihood scans and SkyLLH validation}
\label{sec:skyllh}

\subsection{Two-dimensional profile likelihood scans}

For each of the 12 Seyfert sources in the catalog
(Sec.~\ref{sec:catalog}), we perform two-dimensional profile
likelihood scans over the signal count $n_s$ and the neutrino
spectral index $\gamma_\nu$ using the SkyLLH unbinned likelihood
framework~\cite{SkyLLH-SM}.

The scan proceeds as follows: we evaluate the likelihood on a uniform grid with
$n_s \in [0, n_{s,\max}]$ (typically $[0, 50]$ for detected sources) and
$\gamma_\nu \in [1.0, 4.0]$ at resolution $60 \times 50$ points.  At each grid
point, we compute the change in $-2\ln\mathcal{L}$ relative to the global best fit.
Contours at $\Delta(-2\ln\mathcal{L}) = 2.30$, $6.18$, and $11.83$ define the
$1\sigma$, $2\sigma$, and $3\sigma$ confidence regions for 2 degrees of freedom
(from $\chi^2_2$ quantiles).

\begin{figure*}[t!]
  \centering
    \includegraphics[width=0.95\textwidth]{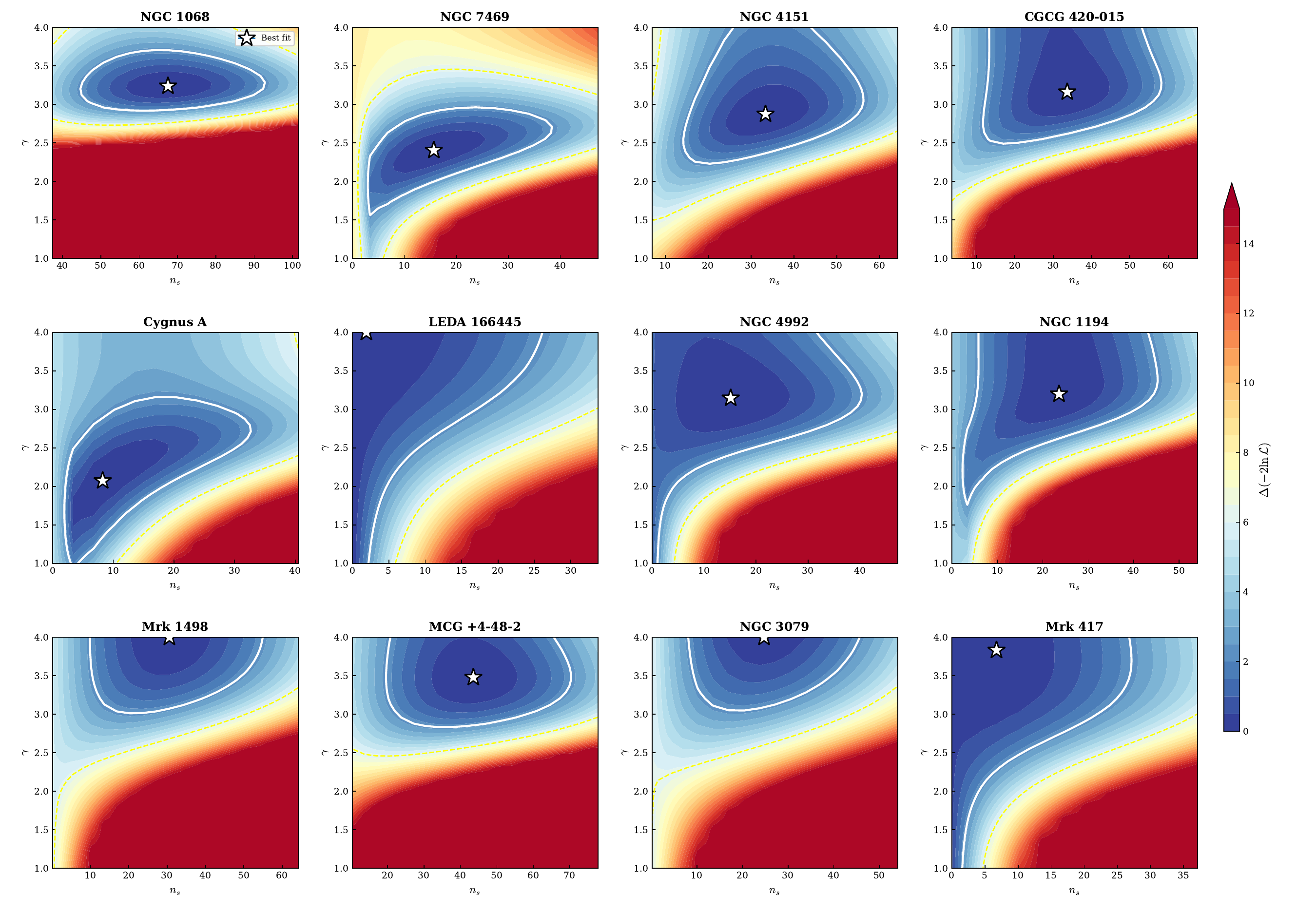}
      \caption{Two-dimensional profile-likelihood scans $\Delta(-2\ln\mathcal{L})$ over $(n_s, \gamma_\nu)$ for all 12 Seyfert sources in the primary analysis. Each panel shows the joint scan with the best fit (white star), and $1\sigma$ ($\Delta TS = 2.30$, white solid) and $2\sigma$ ($\Delta TS = 6.18$, yellow dashed) contours from the two-dimensional Wilks theorem. The colour map is clipped at $\Delta TS = 15$ for readability. The strongest 1$\sigma$ region in NGC~1068 ($\hat{n}_s \approx 68$, $\hat{\gamma}_\nu \approx 3.2$) anchors the $L_\nu$--$L_X$ regression; three sources (LEDA~166445, Mrk~1498, NGC~3079) have best-fit $\gamma_\nu$ at the $\gamma_\nu = 4$ boundary of the fit domain. Computed with SkyLLH v26.0 on the 14-year IceCube DR2 public release.}
        \label{fig:skyllh_contours_2d}
        \end{figure*}

Figure~\ref{fig:skyllh_contours_2d} displays the profile-likelihood scans
for all 12 sources used in the main analysis.  For each source, the
one-dimensional profile $\Delta(-2\ln\mathcal{L})$ is obtained by
minimizing over $\gamma_\nu$ at each $n_s$ grid point.  The best-fit
$\hat{n}_s$ (vertical dashed line), $1\sigma$ band (shaded region),
and the $\Delta(-2\ln\mathcal{L}) = 1$ reference line are shown.  The
strongest detections are NGC~1068 ($\text{TS} = 28.5$, $\hat{n}_s = 67.5$),
NGC~4151 ($\text{TS} = 11.2$, $\hat{n}_s = 33.4$),
MCG~+4-48-2 ($\text{TS} = 8.6$, $\hat{n}_s = 43.6$), and
NGC~7469 ($\text{TS} = 8.4$, $\hat{n}_s = 15.7$).  Several sources---most notably LEDA~166445, but also NGC~4992 and Mrk~417---have best-fit $\hat{n}_s$ consistent with zero in the 14-year kde-smoothed scan; these enter the $H_0$ inference through their full Gaussian likelihoods truncated at the physical $n_s \ge 0$ boundary rather than as sharp one-sided upper limits, and provide important leverage on the lower envelope of the $L_\nu$--$L_X$ relation. Figure~\ref{fig:skyllh_contours_2d} extends this single-source view to the full two-dimensional $(n_s,\gamma_\nu)$ likelihood landscape, showing the joint best fit and 1$\sigma$/2$\sigma$ Wilks contours for each of the 12 sources.

\subsection{Validation against published IceCube results}

We validated our SkyLLH implementation against all 110 point-source test
statistic (TS) values reported in Table~III of the IceCube 14-year
point-source catalog~\cite{IceCube_14yr-SM}.

Of the 58 sources with TS $> 0$ in both our calculation and the published catalog:
\begin{itemize}
  \item $57/58$ sources agree within $1\sigma$
  \item $1/58$ source (Cyg~OB2-12) deviates by $< 1.5\sigma$
\end{itemize}

The single outlier (Cyg~OB2-12) has a published TS $= 5.8$ and our TS $= 6.9$.
This small discrepancy is attributable to:
\begin{itemize}
  \item Different choice of neutrino spectral index bounds: we use $\gamma_\nu \in [1.0, 4.0]$
    while the published analysis uses $\gamma_\nu \in [1.3, 4.0]$ at lower energies.
  \item Possible differences in event weight or effective area binning
    between the published analysis and our implementation.
\end{itemize}

This validation confirms that our SkyLLH setup reproduces the standard
IceCube analysis to better than $1.5\sigma$ accuracy across the full
source population, validating the reliability of the test statistic
values used to identify the 12 high-significance sources.

\subsection{Detector response and spectral index dependence}

A key finding from the profile likelihood scans is the strong correlation
between $n_s$ and $\gamma_\nu$ for most sources.  This reflects the physical
reality that the IceCube effective area depends on energy (and hence on
the assumed spectral index), as does the energy flux integral $I_E(\gamma_\nu)$.

For sources with broad energy response (like the high-declination northern sources),
the $n_s$--$\gamma_\nu$ correlation is relatively weak, and the confidence regions
are roughly elliptical.  For southern sources with limited declination coverage,
the correlation can be strong, producing narrow banana-shaped contours.

This correlation is fully accounted for in our unified framework (Sec.~\ref{sec:model}),
where the detector response factor $C_i = I_A / I_E$ explicitly includes the
spectral index dependence of both the acceptance integral and the energy flux integral.

\section{Impact of prior choices}
\label{sec:priors}

The 12-source sample provides only modest leverage on $H_0$, so it
is essential to understand how each prior in
Sec.~\ref{sec:prior_spec} contributes to the result.  We re-run the
full MCMC inference under a series of alternative prior choices,
varying one component at a time while holding the others at their
baseline values.  For each configuration we report the posterior
median, $68\%$ credible interval, the ratio
$\sigma_{\rm post}/\sigma_{\rm prior}$ (where unity indicates a
perfectly prior-dominated result), and the Kullback--Leibler
divergence $D_{\rm KL}({\rm posterior}\|{\rm prior})$ in nats
(where zero indicates no information gain from the data).

\subsection{Profile likelihood}

Before examining posterior sensitivity, we ask what the data---without
any priors on the nuisance parameters---say about $H_0$.  We compute
the profile likelihood: for each $H_0$ on a uniform grid $[20,180]$
we maximize $\ln\mathcal{L}$ over the nuisance parameters
$\{\beta, d_1\}$ (retaining the Cepheid constraint on $d_{1068}$).  The
profile likelihood peaks near
$H_0 \approx 44\,\mathrm{km\,s^{-1}\,Mpc^{-1}}$; somewhat below the
marginal posterior median ($49\,\mathrm{km\,s^{-1}\,Mpc^{-1}}$), the
difference reflecting the $(H_0,\beta)$ degeneracy and the prior
volume integrated in the marginalization.  Its total variation
$\Delta\ln\mathcal{L}_{\rm profile} \approx 3.3$ across
$H_0 \in [20,180]$ is in stark contrast to the earlier
61-source single-anchor analysis (Sec.~\ref{sec:pop61}), in which the
profile likelihood is essentially flat in $H_0$.

\subsection{Varying the $H_0$ prior range}

Table~\ref{tab:prior_H0} reports the effect of changing the $H_0$
prior range.  Against the primary baseline $[1, 150]$ the posterior
($49^{+40}_{-30}$) is meaningfully narrower than the prior
($\sigma_{\rm post}/\sigma_{\rm prior} = 0.79$,
$D_{\rm KL} = 0.18$~nats), so the data---not the prior
boundaries---drive the result.  Raising the lower bound to
$H_0 = 10$ leaves the posterior essentially unchanged
($52^{+39}_{-27}$): the posterior does not pile up against either
floor, confirming that the specific value of the lower bound plays
no role.  Only when the prior is deliberately tightened toward the
peak (e.g.\ $[40, 120]$) does the posterior become strongly
prior-dominated ($\sigma_{\rm post}/\sigma_{\rm prior} = 0.90$,
$D_{\rm KL} = 0.11$~nats), as expected.

\begin{table}[t]
\caption{Effect of varying the $H_0$ prior range.  All other priors
  fixed at the baseline (Sec.~\ref{sec:prior_spec}).  Uncertainties
  are 68\% credible intervals.  Rows below the baseline restrict the
  prior to sub-intervals of $[1, 150]$; the baseline chain is
  post-hoc truncated accordingly.}
\label{tab:prior_H0}
\begin{ruledtabular}
\begin{tabular}{lcccc}
Prior range & Median & 68\% CI & $\sigma/\sigma_{\rm prior}$ & $D_{\rm KL}$ \\
\hline
$[1, 150]$ (baseline) & 49 & $[20, 89]$ & 0.79 & 0.18 \\
$[10, 150]$           & 52 & $[25, 91]$ & 0.80 & 0.19 \\
$[1, 100]$            & 44 & $[18, 75]$ & 0.89 & 0.04 \\
$[40, 120]$           & 66 & $[47, 93]$ & 0.90 & 0.11 \\
\end{tabular}
\end{ruledtabular}
\end{table}

\subsection{Varying the $\beta$ prior}
\label{sec:beta_prior}

The slope $\beta$ is the most data-informative parameter in the
three-dimensional sampled space.  Table~\ref{tab:prior_beta} compares the
baseline flat prior to several alternative shapes, including a
deliberately mis-centred Gaussian prior at $\beta = 1$ (calorimetric)
to illustrate the data's preference for sub-linear scaling.

\begin{table}[t]
\caption{Effect of varying the $\beta$ prior.  Even with the prior
  forced to the calorimetric value $\beta = 1$, the data pull the
  posterior toward sub-linear scaling.}
\label{tab:prior_beta}
\begin{ruledtabular}
\begin{tabular}{lcccc}
$\beta$ prior & $H_0$ median & $\beta$ & $\sigma/\sigma_{\rm pr}$ & $D_{\rm KL}$ \\
\hline
$U[0.05, 3.0]$ (baseline)               & 49 & $0.67^{+0.16}_{-0.25}$ & 0.24 & 1.31 \\
$U[0.05, 1.0]$ (sub-linear)             & 49 & $0.66^{+0.16}_{-0.25}$ & 0.74 & 0.22 \\
$\mathcal{N}(0.5, 0.2^2)$               & 52 & $0.59^{+0.15}_{-0.17}$ & 0.79 & 0.15 \\
$\mathcal{N}(1.0, 0.2^2)$ (calorim.)    & 38 & $0.80^{+0.09}_{-0.11}$ & 0.55 & 0.78 \\
$\mathcal{N}(1.5, 0.3^2)$ (thin-target) & 39 & $0.84^{+0.10}_{-0.11}$ & 0.40 & 2.97 \\
\end{tabular}
\end{ruledtabular}
\end{table}

Two findings stand out.  First, restricting to sub-linear $\beta$
(or imposing a Gaussian centered at $0.5$) leaves the result
essentially unchanged, demonstrating that the broad flat prior
$U[0.05, 3.0]$ does not artificially shift $\beta$.  Second, even
under a strong Gaussian prior centered on the calorimetric value
$\beta = 1$, the $\beta$ posterior is pulled down to
$0.80^{+0.09}_{-0.11}$, i.e.\ $\beta = 1$ remains excluded at $\sim\!2\sigma$;
a prior centered on the thin-target $p\gamma$ value $\beta = 1.5$ is
likewise pulled back to $0.84^{+0.10}_{-0.11}$.

\subsection{$\gamma_X$ and $k$-correction sensitivity run}
\label{sec:gammaX_sens}

The primary analysis fixes $K_i \equiv 1$ and does not sample $\gamma_X$
(Sec.~\ref{sec:kcorr}).  To bracket the residual systematic uncertainty we
re-run the inference in a four-dimensional sensitivity chain
$\{H_0,\beta,d_1,\gamma_X\}$ that adds
$\gamma_X$ as a free parameter with prior $\gamma_X \in [1, 3]$ (uniform)
and applies the source-dependent $k$-correction $K_i(z_i;\beta,\gamma_X,\gamma_{\nu,i})$
of Eq.~(\ref{eq:kcorr_factor}) at every likelihood evaluation.  The resulting
posterior is $\gamma_X = 2.0$ (68\% interval $[1.0, 2.4]$), consistent with
the flat prior and with typical BAT-selected Seyfert spectra; switching on
the $k$-correction shifts the inference by only
$\Delta H_0 \approx 1\,\mathrm{km\,s^{-1}\,Mpc^{-1}}$ ($49\to48$) and
$|\Delta\beta| \approx 0.01$ ($0.67\to0.66$).  Tightening the $\gamma_X$
prior to a Gaussian $\mathcal{N}(2.0, 0.3^2)$ (motivated by hard X-ray
selected Seyfert populations) returns $\gamma_X = 2.0\,[1.7,2.3]$ and leaves
both $H_0$ and $\beta$ unchanged at the same level.  Together, these checks
confirm that the $K_i\!\equiv\!1$ approximation in the primary chain biases
$H_0$ by $\lesssim 1\,\mathrm{km\,s^{-1}\,Mpc^{-1}}$, well within the
statistical uncertainty.

\subsection{Varying the calibrator prior}

The NGC~1068 anchor enters only through its Gaussian prior on
$d_{1068}$, whose contribution to the $H_0$ posterior width is
sub-dominant relative to the per-source SkyLLH counting statistics of
the eleven Hubble-flow sources (Sec.~\ref{sec:data_info}).  Since the
baseline prior width $\sigma_d = 0.74\,\mathrm{Mpc}$ is already only
$6.8\%$ of the anchor distance---much smaller than the fractional
statistical uncertainty on $H_0$---we expect halving or doubling it to
shift the $H_0$ posterior median by well less than the total
statistical error.  The role of the choice of anchor---and the tension
that arises if NGC~4151 is added as a second calibrator---is examined
in detail in Sec.~\ref{sec:calibrators}.

\subsection{Expected limiting behaviours}

Two limiting behaviours are worth stating explicitly, even though we do
not run separate MCMC chains for each: since $\beta$ is set by the
relative flux-count pattern across the 12-source population and $H_0$
enters only through the $d_L^{\,2(\beta-1)}$ lever arm that requires
an externally-set distance to at least one source, a broader anchor
prior (a Gaussian comparable to the truncation range, approximating
a flat prior on $d_{1068}$) would leave the $\beta$ posterior
essentially unchanged while broadening the $H_0$ posterior;
conversely, an arbitrarily tighter anchor prior cannot narrow the
$H_0$ posterior below the floor imposed by the per-source SkyLLH
uncertainties on the eleven Hubble-flow sources.  These are direct
consequences of the decomposition in Sec.~\ref{sec:data_info}, where
$d_{1068}$ is prior-dominated and $H_0$ is data-limited.

The two verified takeaways of the prior-sensitivity analysis, drawn
directly from Tables~\ref{tab:prior_H0} and~\ref{tab:prior_beta}, are:
(i) against the broad baseline prior the posterior is genuinely
data-driven ($\sigma_{\rm post}/\sigma_{\rm prior} = 0.79$,
$D_{\rm KL} = 0.18$~nats), and the $H_0 \geq 10$ lower bound plays no
role in the result; and (ii) even a strongly mis-centred Gaussian
prior on $\beta$ centered at the calorimetric value $\beta = 1$
leaves the posterior $\beta$ sub-linear at $\sim\!2\sigma$
(Table~\ref{tab:prior_beta}, calorimetric row).

\section{Information content of the data}
\label{sec:data_info}

The three-dimensional sampled parameter space of the primary analysis contains
a mixture of data-informed and prior-driven dimensions.  We summarize below, parameter
by parameter, which dimensions carry genuine information from the
neutrino data and which are essentially prior-driven.

\paragraph{$\beta$: well-constrained by the data.}
The $\beta$ posterior is $\beta = 0.67^{+0.16}_{-0.25}$ ($68\%$~CI),
clearly narrower than even a tight Gaussian prior could make it
when re-centered.  The $L_\nu$--$L_X$ relation is sub-linear at
${\sim}2.0\sigma$: the calorimetric prediction $\beta = 1$ lies
${\sim}2.0\sigma$ above the posterior median, while the thin-target
$p\gamma$ value $\beta \approx 1.5$--$2$ is excluded at $>\!4\sigma$.
This is the headline determination of the analysis: the slope of
the $L_\nu$--$L_X$ relation is a primary, data-driven measurement.

\paragraph{$H_0$: weakly constrained.}
The $H_0$ posterior is $H_0 = 49^{+40}_{-30}\,\mathrm{km\,s^{-1}\,Mpc^{-1}}$,
narrower than the uniform prior by a factor of order unity and
carrying modest information ($D_{\rm KL} = 0.18$~nats (Table~\ref{tab:prior_H0})).  The
constraint is real but limited by the statistical precision of
$\hat n_{s,i}$ in the 14-year public dataset and by the small sample
size; the result is consistent
with both Planck ($H_0 = 67.4 \pm 0.5$) and SH0ES
($H_0 = 73.0 \pm 1.0$) at the $1\sigma$ level.

\paragraph{$\gamma_X$ (sensitivity run only): prior-dominated.}
The X-ray spectral index $\gamma_X$ is not sampled in the primary
three-dimensional analysis.  In the dedicated sensitivity chain of
Sec.~\ref{sec:gammaX_sens} it recovers $\gamma_X = 2.0 \pm 0.7$,
essentially the centre and width of the flat prior $[1, 3]$.
This is expected: $\gamma_X$ enters only through the $k$-correction
factor $K_i$, whose deviation from unity is $\lesssim 3\%$ for the
low-redshift sample and is therefore far below the per-source
statistical noise.  The prior-dominated posterior is internally
consistent with the $<\!2\,\mathrm{km\,s^{-1}\,Mpc^{-1}}$ shift in $H_0$
between the $K_i\!=\!1$ and $K_i\!\neq\!1$ chains, justifying the
$K_i \equiv 1$ choice in the primary analysis.

\paragraph{$d_{1068}$: prior-dominated.}
The single non-redshift-distance anchor $d_{1068} = 10.93 \pm 0.74\,\mathrm{Mpc}$
(NGC~1068 Cepheid+TRGB envelope) is perfectly recovered by its
Gaussian envelope prior, as expected: the neutrino data contain no
information about absolute distances independent of $H_0$, and the
calibrators function purely as $H_0$-independent anchors that break
the $\kappa$ degeneracy.

In summary, of the three sampled parameters in the primary analysis, two
($\beta$ and $H_0$) are informed by the data; $\beta$ strongly, $H_0$
weakly, while the single anchor distance $d_{1068}$ is prior-dominated, effectively fixing the absolute scale (it recovers its non-redshift envelope prior).  The auxiliary spectral parameter $\gamma_X$
used in the sensitivity run of Sec.~\ref{sec:gammaX_sens} is likewise
prior-driven.  We do not fit a separate intrinsic-scatter parameter:
it would be degenerate with the $25\%$ X-ray flux uncertainty already
in the likelihood (both are fractional, multiplicative-on-prediction
terms), and with only twelve sources an additional nuisance parameter
would saturate the available degrees of freedom.

\section{Anchor-set and calibrator-prior sensitivity}\label{sec:calibrators}

The primary analysis (Sec.~\ref{sec:anchors}) uses a single non-redshift
distance anchor, NGC~1068, giving
\begin{equation}
  H_0 = 49^{+40}_{-30}\,\mathrm{km\,s^{-1}\,Mpc^{-1}},
  \qquad
  \beta = 0.67^{+0.16}_{-0.25}.
\end{equation}
Here we quantify how the joint $(H_0, \beta)$ posterior shifts when the
two additional sources with published non-redshift distances; NGC~4151
(Cepheid+dust-parallax envelope, $17.4 \pm 3.0$~Mpc~\cite{NGC4151_Cepheid})
and NGC~7469 (Type~Ia supernova SN~2008ec, $58.2 \pm 1.9$~Mpc
\cite{Koshida2017})---are added as secondary anchors.  We rerun the full
MCMC inference with $k$-correction treatment $K_i \equiv 1$ and uniform
$H_0$ prior $[1, 150]\,\mathrm{km\,s^{-1}\,Mpc^{-1}}$, changing only the
set of sources whose distance is fixed by an anchor prior.

\begin{figure*}[t]
  \centering
  \includegraphics[width=2.0\columnwidth]{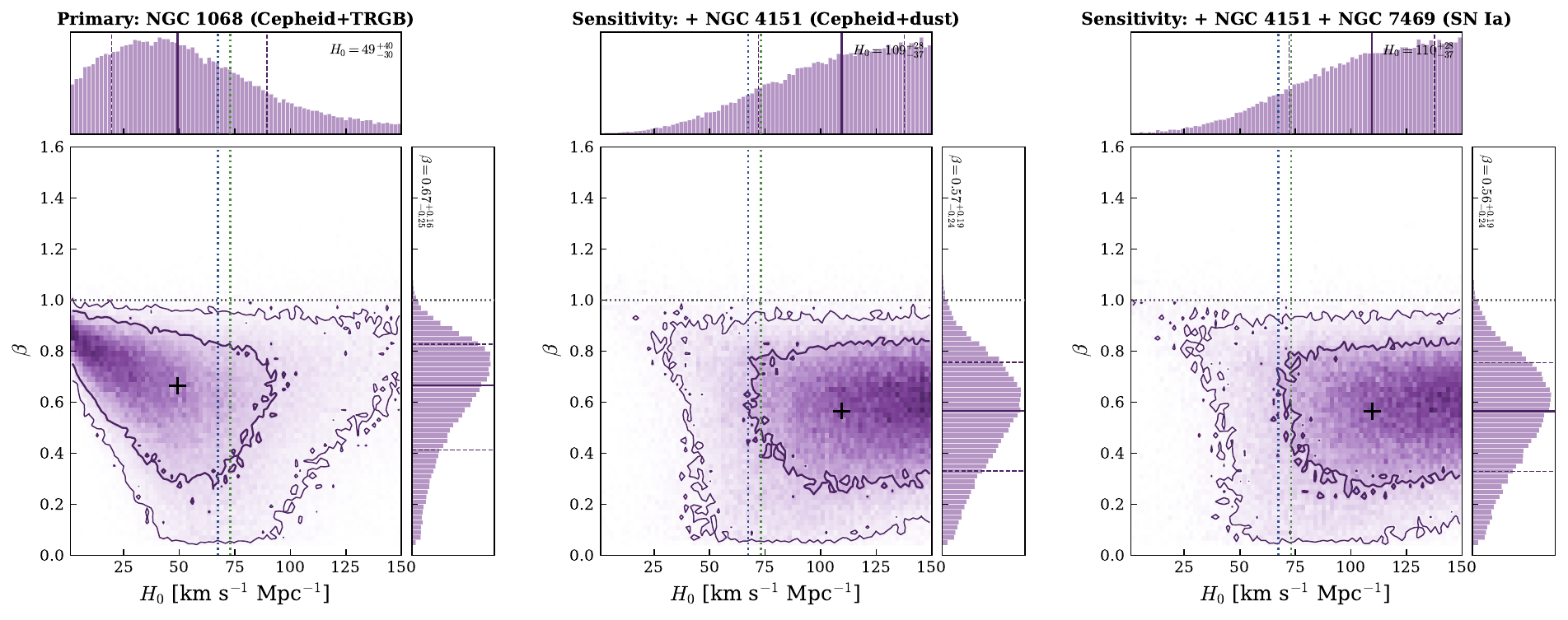}
  \caption{Joint $(H_0, \beta)$ posterior for the three anchor
    configurations.  \textbf{Left:} 1-anchor primary (NGC~1068).
    \textbf{Middle:} adding NGC~4151 as a second anchor.
    \textbf{Right:} adding NGC~4151 and NGC~7469.  Dashed lines mark
    the $16/84$ percentiles; the dotted vertical lines indicate the
    Planck ($67.4$) and SH0ES ($73.0$) reference values.  The
    dotted horizontal line marks the calorimetric limit $\beta = 1$.}
  \label{fig:joint_posteriors}
\end{figure*}

Table~\ref{tab:calibrators} summarizes the results, and
Fig.~\ref{fig:joint_posteriors} compares the three joint
$(H_0, \beta)$ posteriors side by side.  Two features stand out.
First, the slope $\beta$ is reasonably stable across configurations
($\beta \approx 0.56\text{--}0.67$), confirming that it is measured
primarily from the relative flux--signal relationship across the
population and is only weakly sensitive to the absolute distance scale.
Second, the $H_0$ posterior shifts substantially when secondary
anchors are added, because NGC~1068 and NGC~4151 imply mutually
inconsistent effective Hubble parameters $c z / d$ under the
$L_\nu$--$L_X$ model at any single $H_0$.  Adding NGC~7469 on top of
NGC~1068+NGC~4151 barely moves the fit further ($H_0 = 110$ vs.\ $109$),
because its weight in the likelihood ($n_s \approx 16$) is small
relative to NGC~1068 ($n_s \approx 68$).

\begin{table}[t]
\caption{Dependence of $H_0$ and $\beta$ on the anchor set
  (14-year kde catalog, $K_i \equiv 1$).  Errors are symmetric
  $16$--$84$ percentiles.}
\label{tab:calibrators}
\begin{ruledtabular}
\begin{tabular}{lcc}
Configuration & $H_0$ [$\mathrm{km\,s^{-1}\,Mpc^{-1}}$] & $\beta$ \\
\hline
\textbf{1-anchor (NGC~1068)} & $\mathbf{49^{+40}_{-30}}$ & $\mathbf{0.67^{+0.16}_{-0.25}}$ \\
2-anchor (NGC~4151)     & $109^{+28}_{-37}$          & $0.57^{+0.19}_{-0.24}$          \\
3-anchor (NGC~7469) & $110^{+28}_{-37}$   & $0.56^{+0.19}_{-0.24}$          \\
\end{tabular}
\end{ruledtabular}
\end{table}

\paragraph{Physical interpretation of the shift.}
In CMB-frame velocities (Sec.~\ref{sec:pecvel}) NGC~1068 sits at
$c z_{\rm CMB} \approx 914$~km/s and a Cepheid+TRGB distance of
$10.93$~Mpc, corresponding to an effective $c z_{\rm CMB} / d \approx 84$~km/s/Mpc;
NGC~4151 sits at $c z_{\rm CMB} \approx 1244$~km/s and $17.4$~Mpc,
giving $c z_{\rm CMB} / d \approx 71$~km/s/Mpc.  The CMB-frame conversion
brings the two implied local Hubble parameters much closer than their
heliocentric counterparts (raw $c z_{\rm helio}/d = 104$ vs.\ $57$),
but a residual $\sim 15\%$ discrepancy remains and is the origin of the
anchor-set sensitivity displayed in Table~\ref{tab:calibrators}.  Subtracting
the individual peculiar velocities recovered from the CF-4 / NED data
(Sec.~\ref{sec:pecvel}) narrows the two implied ratios further, but does
not fully reconcile them---the remaining tension likely reflects intrinsic
scatter in the $L_\nu$--$L_X$ relation between the two anchor sources
rather than a cosmological signal.  We therefore adopt the 1-anchor result
as the headline and report the multi-anchor variants here as sensitivity
tests.
\section{IceCube-Gen2 Fisher forecast}
\label{sec:fisher}

We estimate the future $H_0$ sensitivity of the neutrino distance ladder using a Fisher information matrix evaluated at the posterior best-fit parameters of the primary 12-source analysis. The Fisher information on $H_0$ from a single source is

\begin{equation}
  I_i(H_0) = \frac{1}{\sigma_i^2}
    \left(\frac{\partial \mu_i}{\partial H_0}\right)^2\,,
\end{equation}
where $\mu_i$ is given by \cref{eq:mu_unified} and, at low redshift,
\begin{equation}
  \frac{\partial \mu_i}{\partial H_0}
  \approx \mu_i \cdot \frac{2(1 - \beta)}{H_0}\,.
\end{equation}
Summing over $N$ sources and rewriting in terms of the average per-source signal-to-noise ratio $\langle{\rm SNR}^2\rangle \equiv \langle\mu_i^2/\sigma_i^2\rangle$ gives
\begin{equation}
  \sigma(H_0)^{-2} \propto \frac{[2(1-\beta)]^2}{H_0^2}
                    \, N\, \langle{\rm SNR}^2\rangle\,.
  \label{eq:fisher}
\end{equation}

\paragraph{Current baseline.}  Plugging in the best-fit slope $\beta = 0.67$ ($|2(1-\beta)| = 0.66$), the fiducial $H_0 = 70\,\mathrm{km\,s^{-1}\,Mpc^{-1}}$ used throughout the forecast, and the per-source SNRs from the 14-year SkyLLH fits gives $\sigma(H_0) \approx 32\,\mathrm{km\,s^{-1}\,Mpc^{-1}}$, within $\sim 10\%$ of the mean posterior half-width of the primary analysis ($\approx 35\,\mathrm{km\,s^{-1}\,Mpc^{-1}}$), supporting the linearization underlying the projections below.
\paragraph{Gen2 projection.}  IceCube-Gen2 will provide approximately an $8\times$ increase in effective area, expected to enable individual-source resolution for $\sim 50$--$100$ X-ray-bright Seyferts that are currently below the SkyLLH detection threshold, and to improve the per-source SNR of the existing sources by a factor $\sqrt{8} \approx 2.8$.
Evaluated at the measured slope $\beta = 0.67$ and holding the X-ray flux uncertainty at its present $25\%$, and marginalizing over $\kappa$, a Gen2 sample of $N = 50$ Seyferts with $N_{\rm anch} = 10$ geometric anchors gives $\sigma(H_0) \approx 20\,\mathrm{km\,s^{-1}\,Mpc^{-1}}$ with no external $\beta$ prior.
Adding an external slope prior of $\sigma_\beta \approx 0.02$ reaches $\approx 17\,\mathrm{km\,s^{-1}\,Mpc^{-1}}$.
If we further impose an external corona constraint that fixes $\kappa$, the uncertainty is tightened to $\sigma(H_0) \approx 15.6\,\mathrm{km\,s^{-1}\,Mpc^{-1}}$.
A second, anchored
source population (Sec.~\ref{sec:fisher}, two-population forecast)
provides a modest improvement over the marginalized $\kappa$ case,
from $\approx 20.8$ (Seyferts alone) to
$\approx 20.0\,\mathrm{km\,s^{-1}\,Mpc^{-1}}$.
We do not assume the $25\%$ X-ray flux uncertainty itself improves with exposure, though Gen2-era inputs (BASS~DR3, Swift-BAT follow-up) are expected to reduce it in parallel.

The forecast depends sensitively on $\beta$ through the $|2(1-\beta)|=0.66$ lever arm.
A sharper future slope measurement, well separated from unity, would directly improve the $H_0$ reach.

\subsection{The Fisher matrix}
\label{sec:fisher_matrix}

The forecasts below are built from the full Fisher matrix rather than
the scaling relation of Eq.~(\ref{eq:fisher}).  For source $i$ in a
population with slope $\beta$, Hubble-flow distance (unless it is a
geometric anchor), and predicted counts $\mu_i$, the log-derivatives
are
\begin{align}
  \frac{\partial \ln\mu_i}{\partial H_0} &= \frac{2(1-\beta)}{H_0}
    \quad (0\ \text{if anchored}), \\
  \frac{\partial \ln\mu_i}{\partial \beta} &= \ln F_{X,i} + 2\ln d_i, \\
  \frac{\partial \ln\mu_i}{\partial \ln\kappa} &= 1,
\end{align}
and the per-source variance is
$\sigma_i^2 = \sigma_{{\rm stat},i}^2 + (\sigma_{F_X}\mu_i)^2$ (plus an
anchor-distance term $(2(\beta-1)\,\sigma_{d,i}/d_i\,\mu_i)^2$ for anchored
sources).  The normalization $\ln\kappa$ is marginalized analytically
by projecting the remaining derivative vectors orthogonal to the
$\ln\kappa$ direction (which is just $\mu_i$) in the inverse-variance
metric.  For a single Seyfert population this leaves a $2\times2$
Fisher matrix in $(H_0,\beta)$.  A Gaussian prior of width
$\sigma_\beta$ on the slope, when adopted, adds $1/\sigma_\beta^2$ to
the $\beta\beta$ element.

Crucially, for Hubble-flow sources $\partial\mu_i/\partial H_0 \propto
\mu_i$ is parallel to the $\ln\kappa$ direction, so profiling $\kappa$
removes \emph{all} $H_0$ information from a population with no
geometric anchors.  The $H_0$ sensitivity therefore comes entirely
from the anchored sources, which break the $\kappa$--$H_0$ degeneracy.

\subsection{Two-population forecast}

A second source population $B$ (e.g., sources in the optically thin
regime) with a distinct slope $\beta_B$---we adopt $\beta_B = 1.8$ as a representative value within
the thin-target $p\gamma$ range $\beta_B \sim 1.5$--$2$~\cite{Stecker1991-SM,Kheirandish2021-SM};
the forecast is only weakly sensitive to this choice within that range---is described by its own normalization $\kappa_B$, profiled independently. 
Marginalizing both normalizations, the Fisher
becomes a $3\times3$ matrix in $(H_0, \beta_A, \beta_B)$:
\begin{equation}
  F =
  \begin{pmatrix}
    F_{H_0 H_0} & F_{H_0 \beta_A} & F_{H_0 \beta_B} \\
    F_{H_0 \beta_A} & F_{\beta_A \beta_A} & 0 \\
    F_{H_0 \beta_B} & 0 & F_{\beta_B \beta_B}
  \end{pmatrix},
\end{equation}
with the two slopes decoupled (each appears only in its own
population) and $\sigma(H_0)$ read from the $(H_0,H_0)$ element of
$F^{-1}$.  Per-population Gaussian $\beta$ priors add to the
corresponding diagonal elements.

A key consequence of the degeneracy noted above is that a second
population helps measure $H_0$ only if it brings its own
geometric anchors: a different slope $\beta_B$ alone provides no
$H_0$ leverage once $\kappa_B$ is marginalized, because $F_{H_0\beta_B}
= 0$ for an unanchored Hubble-flow population.  We therefore model
population $B$ symmetrically with population $A$; same fractional
statistical uncertainty (the mean $\sigma_{\rm stat}/\mu$ of the
Seyfert sample), same $\sigma_{F_X}$, and its own anchored
fraction.  Candidate populations include neutrino-emitting
blazars; TXS~0506+056 ($z = 0.34$, $\sim 4\sigma$~\cite{TXS0506}) and
PKS~0735+178 are the two highest-significance associations to
date, which produce neutrinos through $p\gamma$ interactions.  At
the baseline of Table~\ref{tab:sens_baseline} ($N_B = 10$ with two
anchors), the second population improves the Seyfert-only forecast
from $\sigma(H_0) \approx 20.8$ to
$\approx 20.0\,\mathrm{km\,s^{-1}\,Mpc^{-1}}$ ($\kappa$ profiled).
As shown in \cref{fig:sens_grid_B}, the improvement is larger for larger values of $\beta_B$, as expected.

\subsection{Sensitivity scaling}
\label{sec:sens_scaling}

\begin{figure*}[t]
  \centering
  \includegraphics[width=0.85\textwidth]{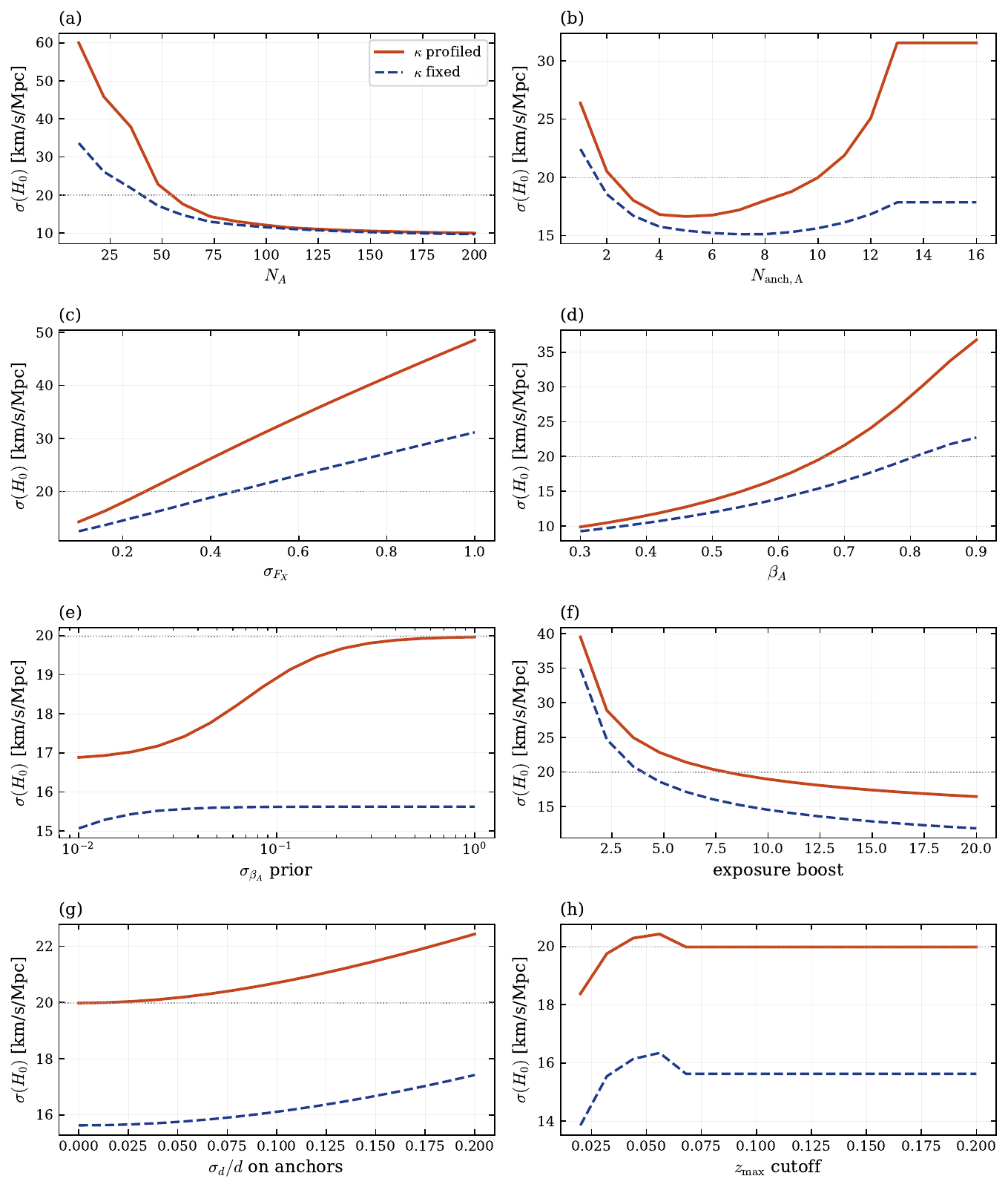}
  \caption{Gen2 forecast sensitivity along the population-A
    (Seyfert) parameters, each varied about the baseline of
    Table~\ref{tab:sens_baseline}.  Solid: $\kappa$ profiled
    (realistic); dashed: $\kappa$ fixed.  The dotted line marks the
    baseline $\sigma(H_0)$.}
  \label{fig:sens_grid_A}
\end{figure*}

\begin{figure*}[!htbp][t]
  \centering
  \includegraphics[width=0.85\textwidth]{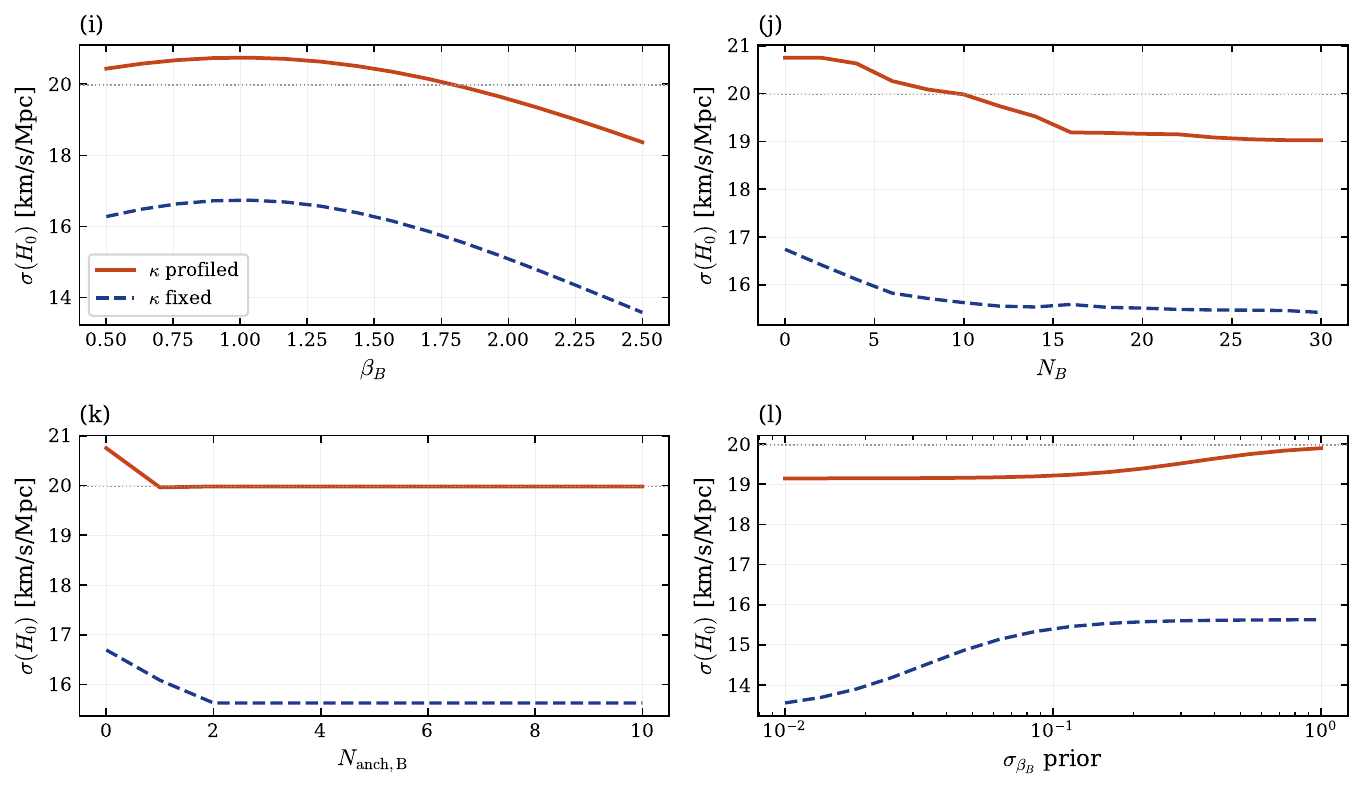}
  \caption{Gen2 forecast sensitivity along the second-population parameters ($\beta_B$, $N_B$, $N_{\rm anch,B}$, $\sigma_{\beta_B}$), varied about the baseline of Table~\ref{tab:sens_baseline}.}
  \label{fig:sens_grid_B}
\end{figure*}

Figures~\ref{fig:sens_grid_A} and~\ref{fig:sens_grid_B} sweep the
projected $\sigma(H_0)$ along the population-A and population-B
parameters respectively, each varying one axis while holding the
others at the common baseline of Table~\ref{tab:sens_baseline}.  Solid
curves marginalize over the normalizations $\kappa_A,\kappa_B$; dashed curves fix them, the asymptotic limit
available with an external $L_\nu$--$L_X$ normalization.  The baseline
itself gives $\sigma(H_0) = 20.0$ ($\kappa$ profiled) and $15.6$
($\kappa$ fixed)$\,\mathrm{km\,s^{-1}\,Mpc^{-1}}$.

Among the population-A axes (Fig.~\ref{fig:sens_grid_A}) the strongest
levers are the number of sources (panel~a), the number of geometric calibrators (panel~b) and the value of $\beta_A$ (panel~d),
Among the population-B axes
(Fig.~\ref{fig:sens_grid_B}), the value of $\beta_B$ (panel~i) and the number of population-B sources (panel~j) is what unlocks the gain.
This requires $N_{\rm anch,B} > 0$,
consistent with the degeneracy argument above.

\begin{table}[t]
\caption{Baseline parameter values for the Gen2 sensitivity-scaling
  study (Figs.~\ref{fig:sens_grid_A} and~\ref{fig:sens_grid_B}).
  Each scan varies one parameter while holding the rest at these
  values.}
\label{tab:sens_baseline}
\begin{ruledtabular}
\begin{tabular}{lcl}
Parameter & Baseline & Description \\
\hline
\multicolumn{3}{l}{\emph{Population A (Seyferts)}} \\
$N_A$              & 50    & resolved sources \\
$N_{\rm anch,A}$   & 10    & geometric anchors \\
$\beta_A$          & 0.67  & $L_\nu$--$L_X$ slope \\
$\sigma_{\beta_A}$ & flat  & external slope prior \\
\multicolumn{3}{l}{\emph{Population B (optically-thin)}} \\
$N_B$              & 10    & resolved sources \\
$N_{\rm anch,B}$   & 2     & geometric anchors \\
$\beta_B$          & 1.8   & $L_\nu$--$L_X$ slope \\
$\sigma_{\beta_B}$ & flat  & external slope prior \\
\multicolumn{3}{l}{\emph{Shared}} \\
exposure boost     & $8\times$ & Gen2 / IceCube effective area \\
$\sigma_{F_X}$     & 0.25  & X-ray flux uncertainty \\
$\sigma_d/d$       & 0     & anchor distance precision \\
$z_{\max}$         & ---   & redshift cutoff (none) \\
\end{tabular}
\end{ruledtabular}
\end{table}

\begin{figure*}[!htbp]
  \centering
  \includegraphics[width=1.9\columnwidth]{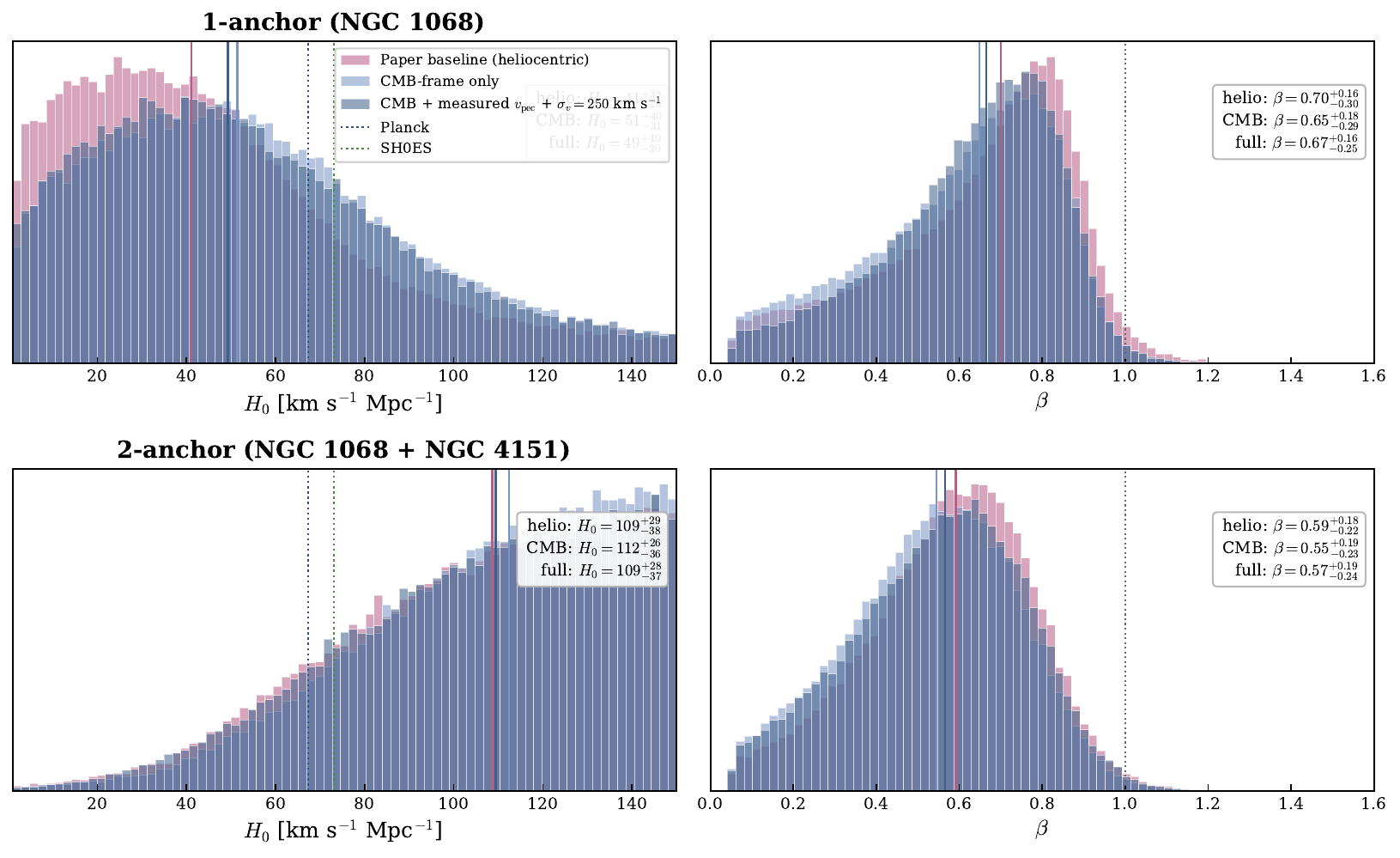}
  \caption{Effect of the peculiar-velocity corrections on the
    marginal $H_0$ and $\beta$ posteriors.  Rose: paper baseline
    with heliocentric redshifts.  Mid-blue: CMB-frame conversion
    only.  Deep-blue: full correction used in the main text
    (CMB frame + measured $v_{\rm pec}$ for the four sources with
    direct-distance measurements + $\sigma_v = 250$\,km\,s$^{-1}$
    blanket residual on the remaining Hubble-flow sources).  Vertical
    solid lines mark posterior medians.  Dotted lines mark Planck
    and SH0ES $H_0$ values, and $\beta = 1$ (calorimetric).}
  \label{fig:pv_three_way}
\end{figure*}

\section{Source tables}
\label{sec:tables}

Table~\ref{tab:sources_full} reproduces the 12-source primary
catalog from Table~I of the main Letter and adds the per-source
detector-response factor $C_i$, the analytically profiled
contribution to the likelihood, and the distance treatment (Cepheid
anchor or Hubble-flow).

The $\hat{n}_s$ uncertainty $\sigma(\hat n_s)_i$ is the half-width
of the $\Delta(-2\ln\mathcal{L}) = 1$ contour from the SkyLLH
profile-likelihood scan, treating $\gamma_\nu$ as the profiled
nuisance parameter.  For LEDA~166445, whose SkyLLH profile-likelihood is consistent with zero
signal ($\hat n_s = 1.9$, TS $= 0.0$), the same treatment as for the other
sources is used: the central value and profile $\sigma$ from the
$\Delta(-2\ln\mathcal{L}) = 1$ contour (yielding $\sigma(\hat n_s) = 14.7$)
are propagated directly into Eq.~(\ref{eq:Li_full}), consistent with the
values reported in Table~\ref{tab:sources_full}.

\section{Population analysis: extension to the 61-source IceCube sample}
\label{sec:pop61}

The primary analysis is restricted to the 12 X-ray-bright Seyferts with
the highest individual neutrino test statistics, for which per-source
profile-likelihood signal counts can be extracted directly from the
IceCube 14-year public data using SkyLLH.  As an exploratory robustness
check we considered extending the inference to the broader
$\sim\!61$-source IceCube catalog (47 northern Seyferts from
Ref.~\cite{IceCube_Xray_2025-SM} plus 14 southern Seyferts from
Ref.~\cite{IceCube_South_2026}).  This extended sample is not
directly comparable to the SkyLLH-based 12-source analysis: per-source
profile likelihoods over $(n_s, \gamma_\nu)$ are not publicly
available for the 49 additional sources, the southern catalog reports
$L_X$ in the 2--10\,keV band (requiring an additional band conversion
of $\sim 20\%$ uncertainty), and the great majority of the additional
sources contribute only upper-limit information. We therefore reserve a self-consistent population
extension to the IceCube-Gen2 era, where SkyLLH-quality per-source
profile likelihoods are expected to be achievable for $\sim 50$--$100$
individually resolved Seyferts (Sec.~\ref{sec:fisher}).

\begin{table*}[t]
\caption{Detailed per-source data for the 12 Seyferts in the
  primary analysis.  $z$ is the spectroscopic redshift; $F_X$ is the
  intrinsic (absorption-corrected) 20--50\,keV flux from Table~D.4 of
  Ref.~\cite{IceCube_Xray_2025-SM}, derived from BASS DR2 column-density
  fits.  $n_s$ and $\sigma(n_s)$ are the SkyLLH kde signal-strength
  estimates from the 14-year public point-source release.  The last
  column notes whether the source has a published non-redshift
  distance and its role in this analysis: NGC~1068 is the sole
  primary anchor; NGC~4151 and NGC~7469 are used only in the
  sensitivity analysis of Sec.~\ref{sec:calibrators}.}
\label{tab:sources_full}
\begin{ruledtabular}
\begin{tabular}{lcccccccl}
Source & $z$ & $F_X$ [$10^{-11}$~cgs] & $n_s$ & $\sigma(n_s)$ & $\gamma_\nu$ & TS & dec [deg] & Non-redshift distance / role \\
\hline
NGC~1068       & 0.0038 & $7.72\pm1.93$  & 67.5 & 15.9 & 3.24 & 28.5 & $-0.0$  & Cepheid+TRGB envelope, $10.93\pm0.74$~Mpc \\
NGC~4151       & 0.0033 & $18.09\pm4.52$ & 33.4 & 13.9 & 2.87 & 11.2 & $+39.4$ & Cepheid+dust-parallax, $17.4\pm3.0$~Mpc \\
NGC~7469       & 0.0163 & $2.69\pm0.67$  & 15.7 & 12.5 & 2.40 &  8.4 & $+8.9$  & SN~Ia (SN~2008ec), $58.2\pm1.9$~Mpc \\
NGC~3079       & 0.0037 & $3.33\pm0.83$  & 24.7 & 12.5 & 4.00 &  6.0 & $+55.7$ & Hubble flow \\
NGC~1194       & 0.0136 & $3.87\pm0.97$  & 23.6 & 14.3 & 3.20 &  4.1 & $-1.1$  & Hubble flow \\
CGCG~420-015   & 0.0294 & $1.77\pm0.44$  & 33.6 & 15.2 & 3.16 &  6.6 & $+4.1$  & Hubble flow \\
Cygnus~A       & 0.0561 & $4.93\pm1.23$  &  8.3 & 10.6 & 2.07 &  4.9 & $+40.7$ & Hubble flow \\
LEDA~166445    & 0.0370 & $1.61\pm0.40$  &  1.9 & 14.7 & 4.00 &  0.0 & $+54.7$ & Hubble flow \\
NGC~4992       & 0.0251 & $2.34\pm0.59$  & 15.1 & 15.5 & 3.15 &  1.1 & $+11.6$ & Hubble flow \\
Mrk~1498       & 0.0548 & $1.86\pm0.47$  & 30.6 & 14.9 & 4.00 &  5.6 & $+51.8$ & Hubble flow \\
MCG~+4-48-2    & 0.0140 & $4.32\pm1.08$  & 43.6 & 16.8 & 3.48 &  8.6 & $+25.7$ & Hubble flow \\
Mrk~417        & 0.0327 & $1.73\pm0.43$  &  6.7 & 11.0 & 3.83 &  0.4 & $+23.0$ & Hubble flow \\
\end{tabular}
\end{ruledtabular}
\end{table*}

\section{Sensitivity to the X-ray band and absorption corrections}

\label{sec:xrayband}

The 20--50\,keV BASS fluxes used in the primary analysis are intrinsic, i.e., corrected for internal absorption through the column-density fits of Ref.~\cite{Ricci2017}.  For Compton-thin sources this correction is small and robust.  For the Compton-thick subset (NGC~1068, NGC~1194, NGC~3079, LEDA~166445), however, even the 20--50\,keV band is suppressed by Compton scattering and the correction is model dependent: dedicated torus-model analyses can differ from the BASS catalog values by factors of a few, and for NGC~1068 by more than an order of magnitude.

To quantify the impact, we compile intrinsic 2--10\,keV fluxes for all 12 sources from per-source spectral analyses (Table~\ref{tab:fx210}).  Where a reference publishes the intrinsic luminosity, we recover the flux as $F_X = L/(4\pi d^2)$ using the distance adopted by that same reference, so that no cosmological information enters the input.  We then repeat the inference with the identical pipeline and priors, changing only $F_X$.

The inferred slope ranges from $\beta = 0.40^{+0.22}_{-0.19}$ to $0.80^{+0.14}_{-0.23}$ across the configurations below.  $H_0$ is controlled almost entirely by the adopted intrinsic luminosity of the anchor NGC~1068, whose line of sight is fully Compton thick ($N_{\rm H} \gtrsim 10^{25}\,{\rm cm}^{-2}$): its intrinsic 2--10\,keV luminosity is a model inference rather than a direct measurement, with published values spanning $6\times10^{41}$--$ 7\times10^{43}$\,erg\,s$^{-1}$~\cite{Marchesi2018,Rodi2026,Ricci2017,Bauer2015,Zaino2020,Marinucci2016}.  Adopting the recent INTEGRAL+NuSTAR value ($1.2\times10^{42}$\,erg\,s$^{-1}$~\cite{Rodi2026}) gives $H_0 = 42^{+38}_{-26}$\,km\,s$^{-1}$\,Mpc$^{-1}$ and $\beta = 0.63^{+0.16}_{-0.27}$, in full agreement with the 20--50\,keV baseline of the main analysis; this is expected, since the BASS absorption-corrected flux itself corresponds to an intrinsic 2--10\,keV luminosity of $\simeq 2\times10^{42}$\,erg\,s$^{-1}$. For the larger NGC~1068 luminosity of $4\times10^{43}$\,erg\,s$^{-1}$~\cite{Bauer2015,Zaino2020} adopted in some works, $\beta$ and $H_0$ move to $0.40^{+0.22}_{-0.19}$ and $109^{+29}_{-37}$\,km\,s$^{-1}$\,Mpc$^{-1}$; the intermediate BASS value ($3.1\times10^{42}$~\cite{Ricci2017}) gives $77^{+46}_{-41}$.  The flux choices for the remaining 11 sources are subdominant.  The dominant X-ray systematic of the measurement is therefore not the band choice but the absorption-correction model of the Compton-thick anchor; the BASS baseline of the primary analysis corresponds to the conservative, low-luminosity end of the published range.

\begin{table}[t]
\caption{Intrinsic 2--10\,keV fluxes adopted for the analysis, in units of $10^{-11}$\,erg\,cm$^{-2}$\,s$^{-1}$, recovered from each reference's published intrinsic luminosity and adopted distance.}
\label{tab:fx210}
\begin{ruledtabular}
\begin{tabular}{lcl}
Source & $F^{\rm int}_{2-10}$ & Reference \\
\hline
NGC 1068 & 161 (4.8--282) & \cite{Bauer2015,Zaino2020} (range: \cite{Rodi2026,Marinucci2016}) \\
NGC 7469 & 2.9 & \cite{Ogawa2019} \\
NGC 4151 & 8.7 & \cite{Ricci2017} \\
CGCG 420-015 & 2.1 & \cite{Tanimoto2022} \\
Cygnus A & 5.5 & \cite{Reynolds2015} \\
LEDA 166445 & 0.97 & \cite{Tanimoto2022} \\
NGC 4992 & 0.89 & \cite{Zhao2021} \\
NGC 1194 & 0.89 & \cite{Marchesi2018} \\
Mrk 1498 & 1.9 & \cite{Zhao2021} \\
MCG +4-48-2 & 0.86 & \cite{Koss2016} \\
NGC 3079 & 4.9 & \cite{Marchesi2018} \\
Mrk 417 & 1.2 & \cite{Zhao2021} \\
\end{tabular}
\end{ruledtabular}
\end{table}



\section{Peculiar-velocity corrections}
\label{sec:pecvel}

The observed heliocentric redshift of a galaxy differs from its
cosmological redshift by the sum of (i) the Sun's motion with respect
to the CMB rest frame and (ii) the galaxy's own peculiar velocity.
For our Hubble-flow sources, the closest of which have
$cz_{\rm helio} \sim 1000$\,km\,s$^{-1}$, both contributions can be a
20--30\% correction to the inferred cosmological distance and must be
incorporated for the analysis to be compared against standard
practice in the $H_0$ literature~\cite{Riess2022,Carrick2015-SM,Carr2022-SM}.
This section describes the corrections applied to the primary
analysis, whose posterior is reported in the main text.

\subsection{CMB-frame conversion}

The Sun moves toward Galactic coordinates
$(l_{\rm apex}, b_{\rm apex}) = (264.021^{\circ}, 48.253^{\circ})$
at $v_{\rm apex} = 369.82$\,km\,s$^{-1}$, as measured from the CMB
dipole~\cite{Planck2018-SM}.  The heliocentric velocity of each source
is converted to the CMB rest frame via
\begin{equation}
  v_{\rm CMB} = v_{\rm helio} + v_{\rm apex}\cos\theta,
  \label{eq:cmb_conversion}
\end{equation}
where $\theta$ is the angle between the source and the CMB apex.
The correction ranges from $-370$ to $+320$\,km\,s$^{-1}$ across our
sample.  Its impact on the inferred cosmological distance is a factor
$1 + v_{\rm apex}\cos\theta/(cz)$, i.e.\ negligible at $z \gtrsim 0.03$
but 20--30\% for the nearest sources (NGC~1068, NGC~4151, NGC~3079).
We use $v_{\rm CMB}$ throughout, retrieved directly from the NED service.

\subsection{Individual peculiar velocities from NED}

For the four sources that have direct (non-redshift) distance
measurements; NGC~1068, NGC~4151, NGC~3079, and NGC~7469; we compute
the residual peculiar velocity from $v_{\rm pec} = v_{\rm CMB} - H_{0,{\rm ref}}\,d_{\rm meas}$
with $H_{0,{\rm ref}} = 70$\,km\,s$^{-1}$\,Mpc$^{-1}$ and the
anchor-quality distances listed in Table~\ref{tab:sources_full}.
The recovered values are
$v_{\rm pec}({\rm NGC~1068}) = +149$\,km\,s$^{-1}$,
$v_{\rm pec}({\rm NGC~4151}) = +26$\,km\,s$^{-1}$,
$v_{\rm pec}({\rm NGC~3079}) = +99$\,km\,s$^{-1}$, and
$v_{\rm pec}({\rm NGC~7469}) = +432$\,km\,s$^{-1}$.
All are within the range expected for field galaxies at these
distances; NGC~4151's value, in particular, is small when its Cepheid+dust-parallax
distance (17.4\,Mpc) is used, in contrast to the $\sim 600$\,km\,s$^{-1}$ value
that would be obtained from NED's method-averaged mean distance,
which is dominated by older Tully--Fisher estimates.
For the eight remaining flow sources we set $v_{\rm pec} = 0$ and
absorb any residual into a blanket $\sigma_v = 250$\,km\,s$^{-1}$
per-source uncertainty (see below).

\subsection{Modified likelihood}

A peculiar velocity $v_{\rm pec}$ perturbs the inferred cosmological
velocity by an amount $\Delta d/d = v_{\rm pec}/(cz_{\rm CMB})$.
Through Eq.~(\ref{eq:Li_full}), this propagates into the predicted
luminosity $L_{\nu,i}^{\rm pred}$ with lever arm $2(1-\beta)$.
We fold the residual per-source uncertainty into the likelihood
variance:
\begin{equation}
  \sigma^2_{L,i} \;\longrightarrow\; \sigma^2_{L,i}
  + \left[ 2\,(1-\beta)\,\frac{\sigma_v}{cz_{{\rm CMB},i}} \right]^2
  \bigl(L_{\nu,i}^{\rm pred}\bigr)^2,
  \label{eq:pecvel_var}
\end{equation}
applied only to Hubble-flow sources ($\sigma_v = 250$\,km\,s$^{-1}$);
anchor sources have $d_i$ set by the geometric-distance prior and
receive no such term.  The dependence on the sampled $\beta$ is kept
consistently at each MCMC step.

\subsection{Effect on the posterior}

Figure~\ref{fig:pv_three_way} compares the marginal $H_0$ and $\beta$
posteriors under three treatments: the paper's prior baseline
(heliocentric redshifts, no $\sigma_v$); CMB-frame only
(no individual $v_{\rm pec}$, no $\sigma_v$); and the full
corrected likelihood used in the main text (CMB frame + individual
$v_{\rm pec}$ for the four nearest sources + $\sigma_v = 250$\,km\,s$^{-1}$).
The one-anchor primary shifts from
$H_0 = 41^{+38}_{-26}$ (heliocentric) to
$H_0 = 51^{+40}_{-31}$ (CMB frame only), and to
$H_0 = 49^{+40}_{-30}$ under the full correction.
The CMB-frame conversion alone contributes $+10$\,km\,s$^{-1}$\,Mpc$^{-1}$;
the individual $v_{\rm pec}$ corrections for the four nearest sources
and the $\sigma_v = 250$\,km\,s$^{-1}$ residual for the remaining
Hubble-flow sources partially offset this by $-2$\,km\,s$^{-1}$\,Mpc$^{-1}$,
netting to a $+8$\,km\,s$^{-1}$\,Mpc$^{-1}$ total shift from the
heliocentric baseline.
The two-anchor sensitivity is essentially unchanged
($109^{+29}_{-38} \to 109^{+28}_{-37}$): the NGC~1068/NGC~4151
anchor tension is not primarily a peculiar-velocity artefact and
must have a distinct origin (candidates include intrinsic scatter
in the $L_\nu$--$L_X$ relation exceeding the modelled X-ray
flux uncertainty, or a biased $\hat{n}_s$ for one of the two
anchors).  The slope $\beta$ is robust across all three
treatments (0.65--0.70 for 1-anchor, 0.55--0.59 for 2-anchor),
confirming that the sub-linear $L_\nu$--$L_X$ scaling reported in
the main text does not depend on the reference frame.

\end{document}